\documentclass[sigconf,screen]{acmart}
\usepackage{tikz}
\usepackage{amsmath}

\usepackage[normalem]{ulem} % TO AVOID UNDERSCORING BIBLIOGRAPHY
\usepackage{enumitem}

\usepackage{subfig}
\usepackage{microtype}

 \newcommand*{\RELEASE}{}
\ifdefined\RELEASE
    \newcommand\old[1]{}
    \newcommand\boris[1]{#1}
    \newcommand\dmitrii[1]{{\color{blue}[Dmitrii]: #1}}
    \newcommand\marios[1]{}
    \newcommand\ed[1]{}
    \newcommand\cor[1]{{#1}}
    \newcommand\shep[1]{{#1}}
\else
    \newcommand\old[1]{}
    \newcommand\boris[1]{{\color{brown}[Boris]: #1}}% Comment by Boris.
    \newcommand\dmitrii[1]{{\color{blue}[Dmitrii]: #1}}% Comment by Dmitrii.
    \newcommand\plamen[1]{{\color{cyan}[Plamen]: #1}}% Comment by Plamen.
    \newcommand\marios[1]{{\color{green}[Marios]: #1}}% Comment by Marios
    \newcommand\cor[1]{{#1}}
    \newcommand\shep[1]{{\color{red}[C]: #1}}% Corrected for camera-ready.
\fi

\newcommand{\tnameabs}{\underline{Re}cord-\underline{a}nd-\underline{P}refetch}
\newcommand{\tname}{Record-and-Prefetch}
\newcommand{\tac}{REAP}
\newcommand{\iname}{vHive}

\setcopyright{acmcopyright}
\acmPrice{15.00}
\acmDOI{10.1145/3445814.3446714}
\acmYear{2021}
\copyrightyear{2021}
\acmSubmissionID{asplos21main-p212-p}
\acmISBN{978-1-4503-8317-2/21/04}
\acmConference[ASPLOS '21]{Proceedings of the 26th ACM International Conference on Architectural Support for Programming Languages and Operating Systems}{April 19--23, 2021}{Virtual, USA}
\acmBooktitle{Proceedings of the 26th ACM International Conference on Architectural Support for Programming Languages and Operating Systems (ASPLOS '21), April 19--23, 2021, Virtual, USA}

\begin{document}

\title{
%REAP: Record-and-Prefetch Orchestrator for Serverless Hosts
Benchmarking, Analysis, and Optimization of Serverless Function Snapshots
}

\author{Dmitrii Ustiugov}
\authornote{Corresponding author: dmitrii.ustiugov@ed.ac.uk.}
\affiliation{%
  \institution{University of Edinburgh}
  \country{United Kingdom}
}
%\email{dmitrii.ustiugov@ed.ac.uk}

\author{Plamen Petrov}
\affiliation{%
  \institution{University of Edinburgh}
  \country{United Kingdom}
}
%\email{plamenpp1@gmail.com}

\author{Marios Kogias}
\authornote{This work was done while the author was at EPFL, Switzerland.}
\affiliation{%
 \institution{Microsoft Research}
 \country{United Kingdom}
}
%\email{marios.kogias@epfl.ch}

\author{Edouard Bugnion}
\affiliation{%
  \institution{EPFL}
  \country{Switzerland}
}
%\email{edouard.bugnion@epfl.ch }

\author{Boris Grot}
\affiliation{%
 \institution{University of Edinburgh}
 \country{United Kingdom}
}
%\email{boris.grot@ed.ac.uk}

%\author[$\dag$]{Dmitrii Ustiugov}
%\author[$\dag$]{Plamen Petrov}
%\author[$\ddag$]{Marios Kogias\thanks{The work was done when the author was at EPFL.}}
%\author[$+$]{Edouard Bugnion}
%\author[$\dag$]{Boris Grot}
%\affil[$\dag$]{University of Edinburgh}
%\affil[$\ddag$]{Microsoft Research}
%\affil[$+$]{EPFL}
%\author{Dmitrii Ustiugov \hspace{0.4in} Plamen Petrov \hspace{0.4in} Marios Kogias \hspace{0.4in} Edouard Bugnion \hspace{0.4in} Boris Grot\\
%\textit{University of Edinburgh}
%\vspace{-2.5em}
%}

\begin{abstract}

Serverless computing has seen rapid adoption due to its high scalability and flexible, pay-as-you-go billing model. In serverless, developers structure their services as a collection of functions, sporadically invoked by various events like clicks.
High inter-arrival time variability of function invocations motivates the providers to start new function instances upon each invocation,
leading to significant cold-start delays that degrade user experience. To reduce cold-start latency, the industry has turned to {\em snapshotting}, whereby an image of a fully-booted function is stored on disk, enabling a faster invocation compared to booting a function from scratch.

\cor{This work introduces \iname{}, an open-source framework for serverless experimentation with the goal of enabling researchers to study and innovate across the entire serverless stack. Using \iname{},} we characterize a state-of-the-art snapshot-based serverless infrastructure, based on industry-leading Containerd orchestration framework and Firecracker hypervisor technologies. We find that the execution time of a function started from a snapshot is \cor{95\%} higher, on average, than when the same function is memory-resident.
We show that the high latency is attributable to frequent page faults as the function's state is brought from disk into guest memory one page at a time.
Our analysis further reveals that functions access the same stable working set of pages across different invocations of the same function. By leveraging this insight, we build \tac, a light-weight software mechanism for serverless hosts that records functions' stable working set of guest memory pages and proactively prefetches it from disk into memory. Compared to baseline snapshotting, {\tac} slashes the cold-start delays by \cor{3.7$\times$}, on average.

\end{abstract}



%\vspace{55pt}

%\settopmatter{printfolios=true}

\date{}
\maketitle

\section{Introduction}
\label{sec:intro}

Serverless computing has emerged as the fastest growing cloud service and deployment model of the past few years, \cor{increasing its Compound Annual Growth Rate (CAGR) from 12\% in 2017 to 21\% in 2018~\cite{market2,market1}}.
%%%%%%%%increasing annually by over 21\%~\cite{market1}.
%Serverless attracts cloud service developers by combining low service costs and faster time to market than conventional clouds can offer. 
In serverless, services are decomposed into collections of independent stateless functions that are invoked by events specified by the developer. The number of active functions at any given time is determined by the load on  that specific function, and could range from zero to thousands of concurrently running instances. This scaling happens automatically, on-demand, and is handled by the cloud provider. Thus, the serverless  model combines extreme elasticity with pay-as-you-go billing where customers are charged only for the time spent executing their requests -- a marked departure from conventional virtual machines (VMs) hosted in the cloud, which are billed for their up-time  regardless of usage.

To make  the serverless model profitable, cloud vendors colocate {\em thousands} of independent function instances on a single physical server, thus achieving high server utilization. A high degree of colocation is possible because most functions are invoked relatively infrequently and execute for a very short amount of time. Indeed, a study at Microsoft Azure showed that 90\% of functions are triggered less than once per minute 
%\boris{would be better to go to a higher percentile  -- e.g., less than once per minute} 
and 90\% of the functions execute for less than 10 seconds~\cite{shahrad:serverless}. 

Because of their short execution time, booting a function (i.e., {\em cold start}) has overwhelmingly expensive latency, and can easily dominate the total execution time. Moreover, customers are not billed for the time a function boots, which de-incentivizes the cloud vendor from booting each function from scratch on-demand. Customers also have an incentive to avoid cold starts because of their high impact on latency~\cite{warm_func}. As a result, both cloud vendors and their customers prefer to keep function instances memory-resident (i.e., {\em warm})
%, often for tens of minutes to hours and even indefinitely
~\cite{google:warm_req,azure:functions,warm_func}. 
However, keeping idle function instances alive wastefully occupies precious main memory, which accounts for 40\% of a modern server's typical capital cost~\cite{agache:firecracker}. With serverless providers instantiating thousands of function on a single server~\cite{agache:firecracker,firecracker:demo}, the memory footprint of keeping all instances warm can reach into hundreds of GBs.  
%For example, conservatively assuming a 100MB footprint of each function and four thousand memory-resident functions sharing a server, as shown in AWS demonstration~\cite{firecracker:demo}, the required memory footprint is 400GB. Reducing this footprint by sharing memory across functions is eschewed by cloud vendors due to security concerns~\cite{agache:firecracker}.

To avoid keeping thousands of functions warm while also eliding the high latency of cold-booting a function, the industry has embraced {\em snapshotting} as a promising solution. With this approach, once a function instance is fully booted, its complete state is captured and stored on disk. When a new invocation for that function arrives, the orchestrator can rapidly load a new function instance from the corresponding snapshot. Once loaded, the instance can immediately start processing the incoming invocation, thus eliminating the high latency of a cold boot.

\cor{
To facilitate deeper understanding and experimentation with serverless computing, this work introduces {\em vHive}, an open-source framework for serverless experimentation, which enables systems researchers to innovate across the entire serverless stack.\footnote{The code is available at \url{https://github.com/ease-lab/vhive}.}
Existing open-source systems and frameworks are ill-suited for researchers, being either incomplete, focusing only on one of the components, such as a hypervisor~\cite{google:gvisor}, or rely on insufficiently secure container isolation~\cite{apache:openwhisk,fission,fn_project,openlambda,kubeless}.
vHive integrates open-source production-grade components from the leading serverless providers, namely Amazon Firecracker~\cite{agache:firecracker}, Containerd~\cite{containerd:industry}, Kubernetes~\cite{k8s}, and Knative~\cite{knative}, that offer the latest virtualization, snapshotting, and cluster orchestration technologies along with a toolchain for functions deployment and benchmarking.
}

\begin{comment}
The first contribution of this work is {\em vHive}\footnote{The code is available at \url{github.com/ease-lab/vhive}}, a complete open-source serverless infrastructure with all required datacenter- and host-level components for end-to-end serverless deployment and benchmarking as scale \boris{either provide a github link or promise to make it available once paper is published}.
This fills an important gap in existing serverless infrastructure that is either proprietary to cloud vendors or incomplete because ... \boris{very briefly explain and cite. Try to include Catalyzer}.
vHive combines existing components that are de-facto industry standards (e.g.,  Containerd~\cite{containerd:industry} and Firecracker MicroVMs~\cite{agache:firecracker}) with a few custom components in lieu of propriety ones. To facilitate serverless systems and architecture research, we release a complete benchmarking framework that allows to study serverless hosts' throughput and function invocation latencies at various load points. The framework includes an experiment coordinator and a number of measurement clients that drive the function request traffic to highly multi-tenant hosts, each running up to thousands of function instances. % in Firecracker VMs managed by the Containerd orchestrator. 
vHive also includes a set of Dockerized ready-to-run Python-based functions from the FunctionBench benchmark suite~\cite{kim:functionbench,kim:practical}. 

\end{comment}

Using vHive, we study the cold-start latency of functions from the FunctionBench suite~\cite{kim:functionbench,kim:practical}, their memory footprint, and their spatio-temporal locality characteristics when the functions run inside Firecracker MicroVMs~\cite{agache:firecracker} as part of the industry-standard Containerd infrastructure~\cite{fc-ctrd:github,containerd:industry}.
We focus on a state-of-the-art baseline where the function is restored from a snapshot on a local SSD, thus achieving the lowest possible cold-start latency with existing snapshotting technology~\cite{fc:snaps,du:catalyzer}.
%an approach that eliminates the function instance's boot time and function initialization overheads~\cite{du:catalyzer,fc:snaps}.

%?? overheads ... \boris{fill in -- basically want to explain that snapshots are state-of-the-art for low-latency cold start} 
%the functions' cold-start latency, memory footprint characteristics along with spatial and temporal locality of memory accesses. We extend Containerd to support the state-of-the-art Firecracker hypervisor's VM snapshotting feature that relies on restoring the VMM and the emulated devices state from a snapshot file and mapping the guest-physical memory file for lazy paging. 
Based on our analysis, we make three key observations. 
%First, restoring from a snapshot yields a much smaller memory footprint (8-34MB) for a given function than cold-booting the function from scratch (148-256 MB) -- a reduction of 87-96\%. 
\cor{First, restoring from a snapshot yields a much smaller memory footprint (8-99MB) for a given function than cold-booting the function from scratch (148-256 MB) -- a reduction of 61-96\%.}
The reason for the greatly reduced footprint is that only the pages that are actually used by the function are loaded into memory. In contrast, when a function boots from scratch, both the guest OS and the function's user code engage functionality that is never used during serving a function invocation (e.g., function initialization).
%and loading libraries). 
%We show that lazy paging allows to reduce the memory footprint of the function instance that is loaded from a snapshot to mere 8-34MBs, i.e., 4-13\% of the memory footprint of the corresponding freshly booted VM.

Our second observation is that the execution time of a function restored from a snapshot is dominated by serving page faults in the host OS as pages are lazily mapped into the guest memory. The host OS serves these page faults one by one, bringing the pages from the backing file on disk. We find that these file accesses impose a particularly high overhead because the guest accesses lack spatial locality, rendering host OS' disk read-ahead prefetching ineffective. 
Altogether, we find that servicing page faults on the critical path of function execution accounts for 95\% of actual function processing time, on average -- a significant slowdown, compared to executing a function from memory (i.e., ``warm'').
%Altogether, we find that servicing page faults on the critical path of function execution slows down function processing by \cor{95\%}\dmitrii{typo: accounts for 95\% of the cold start delay}, on average, compared to executing a function from memory (i.e., ``warm''). 

%\boris{can you quantify?} \dmitrii{to quantify that here and in \S\ref{sec:characterization} we would need to compare to without flushing the page cache which we didn't really want...} \boris{ok, but then how can you make this claim if you don't have data to back it up?}
%, which the function triggers when accessing the guest memory. 

%The host OS serves these page faults by bringing the pages from the guest-memory backing file on disk. We find that these file accesses impose a particularly high overhead because the guest accesses lack spatial locality, rendering host OS' disk read-ahead prefetching ineffective.

Our last observation is that a given function accesses largely the same set of guest-physical memory pages across multiple invocations of the function. For the studied functions, \cor{97\%, on average} of the memory pages are the same across invocations. 

Leveraging the observations above, we introduce \tname{} (\tac) -- a light-weight software mechanism for serverless hosts that exploits recurrence in the memory working set of functions to reduce cold-start latency. 
Upon the first invocation of a function, REAP records a trace of guest-physical pages and stores the copies of these pages in a small working set file. 
On each subsequent invocation, \tac{} uses the recorded trace to proactively prefetch the entire function working set with a single disk read and eagerly installs it into the guest's memory space. 
\tac{} is implemented entirely in userspace, using the existing Linux user-level page fault handling mechanism~\cite{man:userfaultfd}.
Our evaluation shows that \tac{} eliminates \cor{97\%} of the pages faults, on average, and reduces the cold-start latency of serverless functions by an average of \cor{3.7$\times$}.

We summarize our contributions as following:

\begin{itemize}[leftmargin=*]
    \item \cor{We release vHive, an open-source framework for serverless experimentation, combining production-grade components from the leading serverless providers to enable innovation in serverless systems across their deep and distributed software stack.}
    %\item We introduce vHive, an open source end-to-end serverless infrastructure and benchmarking framework built upon industry-standard Firecracker and Containerd technologies.
    \item Using vHive, we demonstrate that the state-of-the-art approach of starting a function from a snapshot results in low memory utilization but high start-up latency due to lazy page faults and poor locality in SSD accesses. We further observe that the set of pages accessed by a function across invocations recurs. 
%    Studied cold-start latency, throughput, and memory footprint of a production-grade Firecracker-containerd host orchestration framework and a representative FunctionBench suite.
%    \item We identify strong data reuse patterns across invocations of the same function.

    \item We present REAP, a record-and-prefetch mechanism that eagerly installs the set of pages used by a function from a pre-recorded trace. REAP speeds up function cold start time by \cor{3.7$\times$}, on average, without introducing memory overheads or memory sharing across function instances.% as compared to the baseline snapshots.
    \item We implement \tac{} entirely in userspace with minimal changes to the Firecracker hypervisor and no modifications to the kernel. 
    \tac{} is independent of the underlying serverless infrastructure and can be trivially integrated with other serverless frameworks and hypervisors, e.g., Kata Containers~\cite{kata:containers} and gVisor~\cite{google:gvisor}.
    %We are going to release \tac{} code by the time the paper is published.
    
    %on average providing 8-34MB memory footprint per running function instance without sharing memory across instances.
\end{itemize}
\section{Serverless Background}
\label{sec:background}

\subsection{Workload Characteristics and Challenges}
\label{sec:serv_characteristics}

Serverless computing or Function as a Service (FaaS) is an increasingly popular paradigm for developing and deploying online services. In the serverless model, the application functionality is sliced into one or more stateless event-driven jobs (i.e., functions), executed by the FaaS provider. Functions are launched on-demand based on the specified event triggers, such as HTTP requests. All major cloud providers support serverless deployments; examples include Amazon Lambda~\cite{aws:lambdas} and Azure Functions~\cite{azure:functions}. 

A recent study of Azure Functions in production shows that serverless functions are short-running, invoked infrequently, and function invocations are difficult to predict~\cite{shahrad:serverless}. Specifically, the Azure study shows that half of the functions complete within 1 second while $>$90\% of functions have runtime below 10 seconds. Another finding is that functions tend to have small memory footprints: $>$90\% of functions allocate less than 300MB of virtual memory. Lastly, 90\% of functions are invoked less frequently than once per minute, albeit $>$96\% functions are invoked at least once per week.

Given these characteristics of functions, the providers seek to aggressively co-locate thousands of function instances that share physical hosts to increase utilization of the provider's server fleet~\cite{agache:firecracker}. For example, a stated goal for AWS Lambda is deploying 4-8 {\em thousand} instances on a single host~\cite{agache:firecracker,firecracker:demo}.

This high degree of colocation brings several challenges. First, serverless functions run untrusted code provided by untrusted cloud service developers that introduces a challenge for security. 
Second, serverless platforms aim to be general-purpose, supporting functions written in different programming languages for a standard Linux environment. As a result, most serverless providers use virtualization sandboxes that either run a full-blown guest OS~\cite{cloud:hypervisor,kata:containers,agache:firecracker,baidu:kata,opennebula:firework} or emulate a Linux environment by intercepting and handling a sandboxed application's system calls in the hypervisor~\cite{google:gvisor}.

%\boris{I am guessing that the previous sentences are an argument for using full-blown virtualization and not using sfork. I would definitely add the virtualization answer right here; otherwise, this problem is unresolved in this section, and the next section only talks about cold-boot delay, so you're leaving the reader with a major unresolved challenge in their  mind.} 
Another challenge for serverless deployments is that idle function instances occupy server memory.  
%; notably, main memory is a major expense item in datacenter servers, accounting for around 40\% of a modern server's cost~\cite{ustiugov:design,agache:firecracker}. 
To avoid wasting memory capacity, most serverless providers tend to limit the lifetime of function instances to 8-20 minutes after the last invocation due to the sporadic nature of invocations, deallocating instances after a period of inactivity and starting new instances on demand.
Hence, the first invocation after a period of inactivity results in a start-up latency that is commonly referred to as the serverless function \emph{cold-start} delay. In the last few years, high cold-start latencies have become one of the central problems in serverless computing and one of the key metrics for evaluating serverless providers~\cite{cold-start-war,serverless:benchmark}.

\subsection{Hypervisor Specialization for Cold Starts}
\label{sec:cold-start101}

As noted in the previous section, leading serverless vendors, including Amazon Lambda, Azure Functions, Google Cloud Functions, and Alibaba Cloud, choose virtual machines (VMs) as their sandbox technology in order to deliver security and isolation in a multi-tenant environment. 
%VMs let clooud service developers package and run nearly arbitrary code, leveraging the commodity Linux environment. \boris{We already said that functions use VMs in Sec 2.2, so this para can be dramatically shortened and combined with the next one.}
%\boris{This paragraph is a bit long. It's important - because we have to justify snapshots - but it could be tightened imo.}
Although historically virtualization is known to come with significant overheads~\cite{randal:ideal}, recent works in hypervisor specialization, including Firecracker~\cite{agache:firecracker} and Cloud Hypervisor~\cite{cloud:hypervisor}, show that virtual machines can offer competitive performance as compared to native execution (e.g., Docker containers), even for the cold-start delays.

Firecracker is a recently introduced hypervisor with a minimal emulation layer, supporting just a single {\em virtio} network device type and a single block device type, and relying on the host OS for scheduling and resource management~\cite{agache:firecracker}. This light-weight design allows Firecracker to slash VM boot time to 125ms and reduces the hypervisor memory footprint to 3MB~\cite{agache:firecracker,firecracker:demo}.
%Indeed, Firecracker is able to boot from an a priori set up root filesystem in impressive 125ms~\cite{agache:firecracker,firecracker:demo}. 
However, we measure that booting a Firecracker VM within production-grade frameworks, such as Containerd~\cite{containerd:industry} or OpenNebula~\cite{opennebula:firework}, takes 700-1300ms since their booting process is more complex, e.g., it includes mounting an additional virtual block device that contains a containerized function image~\cite{fc-ctrd:github,fc-ctrd:deep-dive}. Finally, the process inside the VM, which receives the function invocation in the form of an RPC, takes up to several seconds to bootstrap before it is able to invoke the user-provided function, which may have its own initialization phase~\cite{du:catalyzer}. Together these delays -- which arise on the critical path of function invocation -- significantly degrade the end-to-end execution time of a function.
%may significantly degrade the performance of a cloud service.

\subsection{VM Snapshots for Function Cold Starts}
\label{sec:snaps101}

%\boris{This paragraph is quite rough. Please polish.}
To reduce cold-start delays, researchers have proposed a number of VM {\em snapshotting} techniques~\cite{fc:snaps,du:catalyzer,google:gvisor}. Snapshotting captures the current state of a VM, including the state of the virtual machine monitor (VMM) and the guest-physical memory contents, and store it as files on disk.
Using snapshots, the host orchestrator (e.g., Containerd~\cite{containerd:industry}) can capture the state of a function instance that has been fully booted and is ready to receive and execute a function invocation. When a request for a function without a running instance but with an existing snapshot arrives, the orchestrator can quickly create a new function instance from the corresponding snapshot. Once loading finishes, this instance is ready to process the incoming request, thus eliminating the high cold-boot latency.

Snapshots are attractive because they require no main memory during the periods of a function's inactivity and reduce cold-start delays. The snapshots of function instances can be stored in local storage (e.g., SSD) or in a remote storage (e.g., disaggregated storage service).

%that together comprise a \emph{snapshot} during the time of function inactivity \boris{the inactivity bit is not clear -- you mean the snapshot is formed when the function is inactive?}. The snapshot file compactly stores the virtual machine monitor (VMM) state, the state of emulated network and block devices, as well as the current state of the whole guest memory. Once created, the snapshot can reside on a local disk or in a disaggregated storage service. 
%so that the snapshotted function instance can be safely torn down, freeing the precious main memory.
%however, to date we are not aware of snapshot-based solutions deployed in production.
%Upon a function invocation, the VM can be rapidly restored from the snapshot to the exact form as at the time the snapshot was taken. 

The state-of-the-art academic work on function snapshotting, Catalyzer~\cite{du:catalyzer}, showed that snapshot-based restoration in the context of gVisor~\cite{google:gvisor} virtualization technology can be performed in 10s-100s of milliseconds.\footnote{
  Here we only consider Catalyzer's "cold-boot" design that does not share memory across instances. We discuss Catalyzer's warm-boot designs in \S\ref{sec:related}.
}
%\boris{you are emphasizing file-based; however, the definition above says that all snapshots are file-based. Thus, this is redundant. I think you want to stress instead that it's a {\em local} snapshot - right?} snapshot~\cite{du:catalyzer}.
To achieve such a short start-up time, Catalyzer minimizes the amount of processing on the critical path of loading a VM from a snapshot. 
\shep{First, Catalyzer stores the minimum amount of snapshot state that is necessary to resume VM execution de-serialized to facilitate VM loading.} 
%ORIGINAL First, Catalyzer loads the minimum amount of state, which is de-serialized and de-compressed off the critical path, that is necessary to resume VM execution. 
After that, Catalyzer maps the plain guest-physical memory file as a file-backed virtual memory region and resumes VM execution.
Crucially, the guest-physical memory of the VM is not populated with memory contents, which reside on disk, when the user code of the function starts running. As a result, each access to a yet-untouched page raises a page fault. %that must be served by the host OS. 
These page faults occur on the critical path of function execution and, as we show in \S\ref{sec:characterization}, significantly increase the runtime cost of a function loaded from a snapshot.

Recently, Firecracker introduced their own open-source snapshotting mechanism that follows the same design principles as Catalyzer, which is proprietary. 
%During a period when a function is idle, its instance's current state can be safely stored on local disk or in a storage service while the instance's resources on the host server -- particularly, the memory the instance occupies -- can be freed.
Similarly to Catalyzer, loading a Firecracker VM from a snapshot is done in two phases. First, the hypervisor process loads the state of the VMM and the emulated devices (that we further refer to as \emph{loading VMM} for brevity) and then maps a plain guest-physical memory for lazy paging~\cite{fc:snaps}.

\section{vHive: an Open-Source Framework for Serverless Experimentation}
\label{sec:infra}
\cor{

To enable a deeper understanding of serverless computing platforms, this paper introduces {\em vHive}, an open-source framework for experimentation with serverless computing. As depicted in Fig.~\ref{fig:vhive_overview}, vHive integrates production-grade components from the leading serverless providers, such as Amazon and Google.

%\boris{I don't find this paragraph useful, except the last sentence. In general, we should aim to keep this section short - 1 page ideally. Instead, you need a 1-sentence lead-in saying "To enable deeper understanding and experimentation with ..., this paper introduces .... Your next sentence should be 'vHive adopts Knative.. (from below)}
%Serverless systems have a datacenter-scale distributed architecture that comprises an extensive number of components and services. This section describes the key components of a modern production system, similar to AWS Lambda~\cite{agache:firecracker}, and introduces vHive,\footnote{vHive is pronounced as /vi$\mathpunct{:}$.haiv/ as in a hive of virtualized functions.} a distributed framework for experimentation with serverless systems, integrating the leading open-source components used in modern production systems by AWS, GCP and other providers. Fig.~\ref{fig:vhive_overview} shows the key components of the serverless infrastructure that vHive integrates.

\begin{figure}[t]
\centering
\includegraphics[width=0.98\columnwidth]{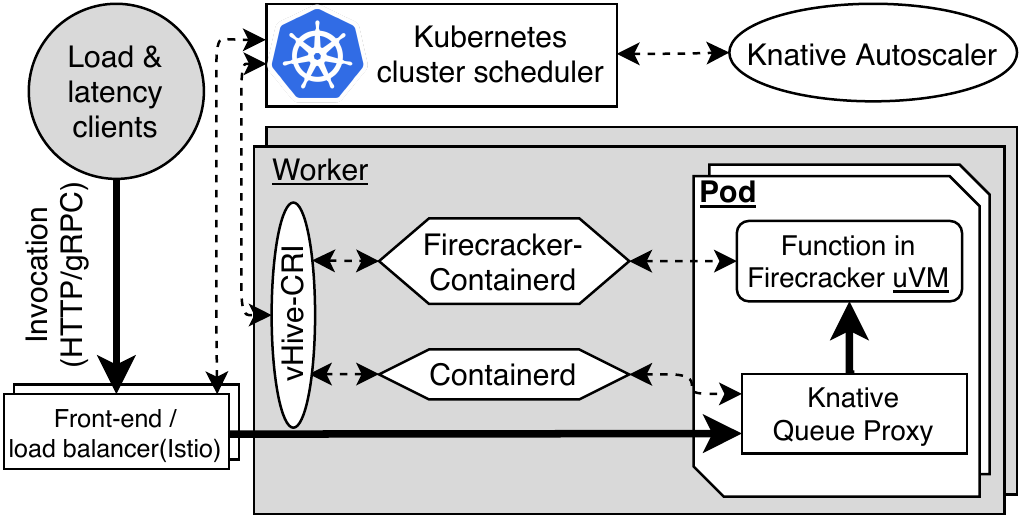}
\caption{vHive architecture overview.
Solid and dashed arrows show the data plane and the control plane, respectively.
}
\label{fig:vhive_overview}
\end{figure}

\subsection{Deploying and Programming with Functions in vHive}
\label{sec:vhive_programming}

%\boris{The following sentence is not useful. You've already established this fact.}
%Serverless platforms allow developers to design their applications as a composition of serverless functions.
%, sometimes using a number of conventional "serverfull" services, e.g., databases, for necessary bookkeeping. 
vHive adopts Knative~\cite{knative}, a serverless framework that runs on top of Kubernetes~\cite{k8s} and offers a programming and deployment model that is similar to AWS Lambda~\cite{aws:lambdas} and Azure functions~\cite{azure:functions}. To deploy an application in vHive, one can deploy application functions by supplying Knative with each function's Open Container Initiative~(OCI)~\cite{opencontainers} image, e.g., a Docker image, along with a configuration file.
% \boris{Does OCI need to be de-acronymized, cited?}
This OCI image contains the function's handle code, which is executed by an HTTP or gRPC server upon an invocation. The configuration file contains the relevant environment variables and other parameters for function composition and function instances scaling.
Using the configuration files, the application developers can compose their functions with any conventional "serverful" services with Kubernetes providing their URLs to the relevant functions. For example, functions that use large inputs or produce large outputs, like photos or videos, often have to save them in an object store or a database.
%\marios{are stateful services relevant here? Also, do stateful services expose functions or just API endpoints? This last sentece is a bit confusing}.

%Upon a function's deployment, Knative provides a URL, using which this function's execution can be triggered\marios{alt: to trigger the function} \boris{I agree w/ Marios. Avoid passive voice!}. 
Upon a function's deployment, Knative provides a URL for triggering this function.
Using these URLs, application developers can compose their functions, e.g., by specifying the URL of a callee function in the configuration file of the caller function. For each function, Knative configures the load-balancer service, sets up the network routes and dynamically scales the number of instances of the function in the system, according to changes in the function's invocation traffic.

\subsection{vHive Infrastructure Components}
\label{sec:vhive_components}

Serverless infrastructure comprises of the front-end fleet of servers that expose the HTTP endpoints for function invocations, the worker fleet that executes the function code, and the cluster manager that is responsible for managing and scaling function instances across the workers~\cite{agache:firecracker,krook:openwhisk,shahrad:serverless}.
These components are connected by an HTTP-level fabric, e.g., gRPC~\cite{google:grpc}, that that enables management and resources monitoring~\cite{agache:firecracker}.

A function invocation, in the form of an HTTP request or an RPC, first arrives at one of the front-end servers for request authentication and mapping to the corresponding function. 
%This front-end server acts as a load balancer for the invocation traffic. 
In vHive, the Istio service~\cite{istio} plays the roles of an HTTP endpoint and a load balancer for the deployed functions.
If the function that received an invocation has at least one active instance, the front-end server simply routes the invocation request to an active instance for processing. 

If there are no active function instances, the load balancer contacts the cluster manager to start a new instance of the function before the load balancer routes this invocation to a worker. vHive relies on Kubernetes cluster orchestrator to automate services deployment and management. 
Knative seamlessly extends Kubernetes, which was originally designed for conventional ``serverful'' services, to enable autoscaling of functions. For each function, Knative deploys an autoscaler service that monitors the invocation traffic to each function and makes decisions on scaling the number of functions instances in the cluster based on observed load.
%} \sout{, from zero to virtually infinity.}
%Over the last decade, Kubernetes has become the framework of choice for cloud providers, including several serverless offerings~\cite{}.

At the autoscaler's decision, a chosen worker's control plane starts a new function instance as a pod, the scaling granule in Kubernetes, that contains a Knative Queue-Proxy~(QP) and a MicroVM that runs the function code. The QP implements a software queue and a health monitor for the function instance, reporting the queue depth to the function's autoscaler, which is the basis for the scaling decisions. The function runs in a MicroVM to isolate the worker host from the untrusted developer-provided code. vHive follows the model of AWS Lambda, which deploys a single function inside a MicroVM that processes a single invocation at a time~\cite{agache:firecracker}.

To implement the control plane, vHive introduces a vHive-CRI orchestrator service that integrates the two forks of Containerd -- the stock version~\cite{containerd:industry} and the Firecracker-specific version developed for MicroVMs~\cite{fc-ctrd:github} -- for managing the lifecycle of containerized services (e.g., the QP) and MicroVMs. The vHive-CRI orchestrator receives Container-Runtime Interface~(CRI)~\cite{cri-o} requests from the Kubernetes control plane and processes them, making the appropriate calls to the corresponding Containerd services. Once the load balancer, which received the function invocation, the QP, and the function instance inside a MicroVM establish the appropriate HTTP-level connections, the data plane of the function is ready to process function invocations. When the function instance finishes processing the invocation, it responds to the load balancer, which forwards the response back to the invoking client.

%%vHive-CRI supports snapshots to facilitate function cold starts, allowing systems researchers to experiment with the cutting edge snapshotting technology in vHive. For generality, vHive also supports starting functions by booting a VM from an image. 

%%%we \boris{the entire sec until this point is written in 3rd person. Please keep it that way and be cognisant of which voice you're using in the future - 3rd voice is generally preferred, but consistency is most important.} extend vHive-CRI and Firecracker Containerd with snapshotting support to facilitate function cold starts,
%, which is at the developer-preview stage in Firecracker roadmap as of December 2020~\cite{firecracker:roadmap} \boris{this info will be outdated within a year. Try not to include such stateements into camera-ready papers that will hopefully persist for a long time}, 
%allowing systems researchers to experiment with the cutting edge snapshotting technology in vHive.
%\footnote{At the moment of publication, no serverless platforms have adopted snapshots in production, to the best of our knowledge.} 
%For generality, vHive also supports starting functions by booting a VM from an image. 

%\subsection{Performance analysis with vHive}

%\boris{Please read my comments throughout this entire section. I think that there's only 1 useful paragraph here, which could be comfortably folded into the prev subsection. Then you don't need subsections at all. }

vHive enables systems researchers to experiment with serverless deployments that are representative of production serverless clouds.
%%%vHive enables systems researchers to set up serverless deployments that are representative of production serverless clouds.\marios{topic sentece is wrong, should be something about experiments, e.g. To help researchers run experiments and collect data,...} \boris{That's a good point, but 'representative' is also important. What about this: "vHive enables systems researchers to experiment with serverless deployments that are representative of production serverless clouds."}
%, by composing the leading open-source production-grade components, like Kubernetes and Firecracker.
%To facilitate adoption by prospective users, vHive is released with a set of functions from FunctionBench~\cite{kim:functionbench,kim:practical} (Table~\ref{tbl:functionbench}). 
%The users can painlessly deploy custom functions or applications of interest, in addition to the provided FunctionBench~\cite{kim:functionbench,kim:practical} (Table~\ref{tbl:functionbench}), into their studies by supplying the corresponding Docker images. 
%With a minimal modification of Knative configuration files, the users can compose and study real-world multi-function applications that may access conventional serverfull services, e.g., databases, deployed in the same Kubernetes cluster.
vHive allows easy analyzing of the performance of an arbitrary serverless setting by offering access to Containerd and Kubernetes logs with high-precision timestamps or by collecting custom metrics.  
vHive also includes the client software to evaluate the response time of the deployed serverless functions in different scenarios, varying the mix of functions and the load. 
%\sout{vHive users can perform these measurements, varying the mix of functions and the load, i.e., the invocation traffic, in their serverless setting.}
Finally, vHive lets the users experiment with several modes for cold function invocations, 
including loading from a snapshot or booting a new VM from a root filesystem.
%\sout{both the cutting-edge Firecracker snapshotting technology and booting a VM from a root filesystem.}
}

\section{Serverless Latency and Memory Footprint Characterization}
\label{sec:characterization}

\cor{
In this section, we use vHive to analyze latency characteristics and memory access patterns of serverless functions, deployed in Firecracker MicroVM instances with snapshot support~\cite{fc:snaps}. 

\subsection{Evaluation Methodology}
\label{sec:char_method}

\begin{comment}

\begin{figure}[t]
\centering
\subfloat[Warm function latency. Note the logarithmic Y-axis.]{
    \includegraphics[width=0.98\columnwidth]{figs/lat_boot.pdf}
    \label{fig:lat_warm}}
\newline
\centering

\subfloat[Cold-start latency breakdown for Firecracker's snapshot load mechanism. \boris{LR-training makes this  graph hard to read. I suggest you  cut off the Yaxis at 1000.} ]{
    \includegraphics[width=0.98\columnwidth]{figs/lat_snap.pdf}
    \label{fig:lat_snap}}
\caption{Warm and cold-start latency of the functions. Note that Y axes are different in (a) and (b).}
\label{fig:lat_char}
\end{figure}

\end{comment}

\begin{figure*}[t]
\centering
\includegraphics[width=\textwidth]{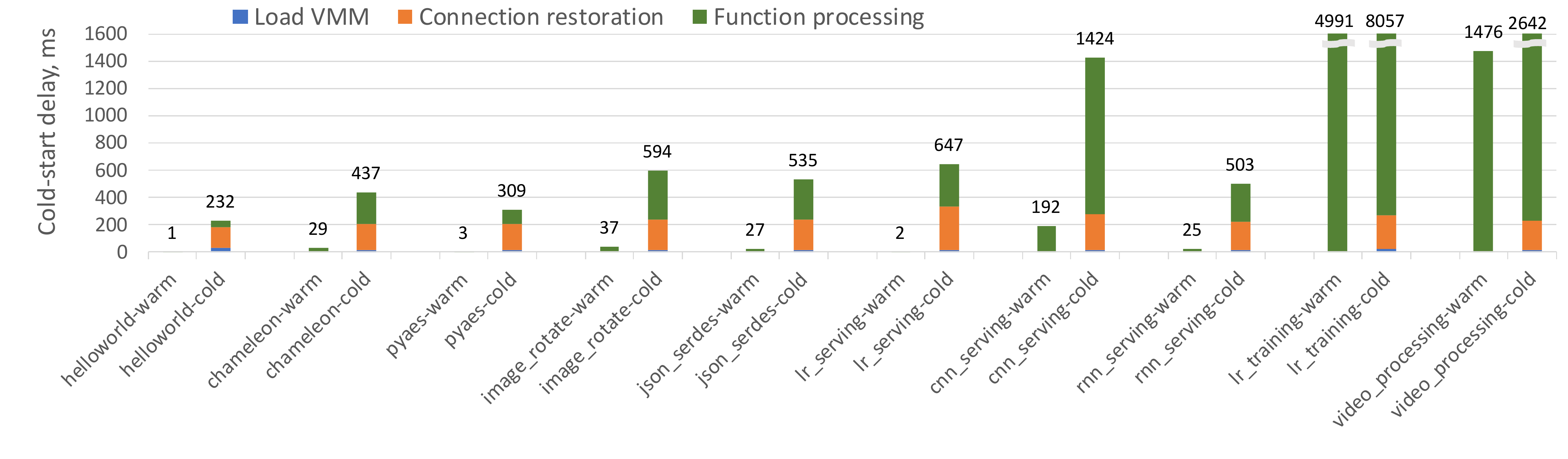}
\vspace{-25pt}
\caption{\cor{Cold-start latency breakdown for Firecracker’s snapshot load mechanism, compared to the warm latency.} 
%\dmitrii{Added recording bar as well} \boris{workload names are messed up/missing. And where is recording?}
}
\label{fig:lat_char}
\end{figure*}

Similarly to prior work~\cite{du:catalyzer}, we focus on the evaluation of a single worker server. The existing distributed serverless stack contributes little, e.g., less than 30ms as shown by AWS~\cite{agache:firecracker}, to the overall end-to-end latency, as compared to many hundreds of milliseconds of the worker-related latency that we demonstrate below. 
Prior work measured the cold-start delay as the time between starting to load a VM from a snapshot to the time the instance executes the first instruction of the user-provided code of the function~\cite{du:catalyzer}. As the metric for our cold-start delay measurements, we choose the latency that includes not only the critical path of the VM restoration but the entire cold function invocation latency on a single worker. 
The measurements capture the latency from the moment a worker receives a function invocation request to the moment when the worker is ready to send the function's response back to the load balancer.
This latency includes both the control-plane delays (including interactions with Containerd and Firecracker hypervisor) and data-plane time that is gRPC request processing and actual function execution.

%For our studies, 
Our experiments aim to closely model the workloads as in a modern serverless environment. First, we adopt a number of functions, listed in Table~\ref{tbl:functionbench}, from a representative serverless benchmark suite called FunctionBench~\cite{kim:functionbench,kim:practical}. 
%Each function is deployed as a handle in a single-thread gRPC server that runs on top of Alpine Linux. 
Second, to simulate the low invocation frequency of serverless functions in production~\cite{shahrad:serverless}, the host OS' page cache is flushed before each invocation of a cold function.

To evaluate the cold-start start delay in a serverless platform similar to AWS Lambda, we augment the vHive-CRI orchestrator
%, further referred to as the vHive orchestrator, 
to act as a MicroManager in AWS~\cite{agache:firecracker}. In this implementation, the vHive-CRI orchestrator not only implements the control plane but also acts as a data plane software router that forwards incoming function invocations to the appropriate function instance and waits for its response over a persistent gRPC connection. Note that in this setting, the worker infrastructure does not include the Queue-Proxy containers so that the data plane resembles
%by\marios{something's wrong with this passive voice, maybe just resembles? emulated?} \boris{Agree. In fact, I am pretty sure you can't say "resembled by" in english. Only 'resembles'.} 
per-function gRPC connections.
%With function snapshotting functionality appearing in just the last year, we are unaware of any studies describing a snapshotting setup in deployment. 
Without a loss of generality, this work assumes the fastest possible storage for the snapshots that is a local SSD, which yields the lowest possible cold-start latency compared to a local HDD or disaggregated storage.
\S\ref{sec:method} provides further details of the platform as well as the host and the guest setup.
%the snapshot files are located on a local SSD, which yields the lowest possible cold-start latency.
%\marios{I would remove the comment about when snapshots were introduced. Just state that you retrieve them from SSDs} \boris{Agree. This may be good for a submission to hammer the novelty point, but for a camera-ready, you need to think about longevity. How is it going to read in 5-10 years?} 
%\marios{Please refer to }\S\ref{sec:method} further details the setup.
%This measurement includes the interval between the invocation is injected by the vHive-CRI orchestrator (called gateway process~\cite{du:catalyzer} or MicroManager in AWS~\cite{agache:firecracker}) directly into a function instance to the point when the orchestrator receives the response from the function instance.
}
%%%%%%As the metric for the cold-start delay measurements, we choose the \emph{end-to-end} latency that captures not only the critical path of the VM restoration but the entire cold function invocation latency, as it is observed by the gateway service of the physical host. This measurement includes the interval between the invocation is injected by the host orchestrator (called gateway process~\cite{du:catalyzer} or MicroManager in AWS~\cite{agache:firecracker}) and the point when the orchestrator receives the response from the function instance. 

%%%%%%%For our studies, we adopt a number of functions from a representative serverless benchmark suite called FunctionBench~\cite{kim:functionbench,kim:practical}. Our evaluation platform is based on our custom-built host orchestrator service that controls the lifecycle of function instances by issuing control requests (e.g., loading an instance from a snapshot) to Containerd~\cite{fc-ctrd:github,fc-ctrd:deep-dive}. The orchestrator also acts as a data plane software router that forwards incoming function invocations to the appropriate function instance and waits for its response over a persistent gRPC connection. With function snapshotting functionality appearing in just the last year, we are not aware of any studies describing a snapshotting setup in deployment. As  such, this work assumes that the snapshot files are located on a local SSD, which yields the lowest possible cold-start latency. \S\ref{sec:method} further details the setup.

To study the memory access patterns of serverless functions, we trace the guest memory addresses that a function instance accesses between the point when the vHive-CRI orchestrator starts to load a VM from a snapshot and the moment when the orchestrator receives a response from the function.  
As Firecracker lazily populates the guest memory, first memory access to each page from the hypervisor or the guest raises a page fault in the host that can be traced. We use Linux \texttt{userfaultfd} feature~\cite{man:userfaultfd} that allows a userspace process to inspect the addresses and serve the page faults on behalf of the host OS.

\begin{comment}
%Our goal is to identify the potential for reducing memory consumption by having cold instances without the cost of cold-start boots.

\boris{mention that we assume local storage for the snapshots, which is the best case latency wise}
To the best of our knowledge, no serverless providers use snapshots in their  serverless production deployments, so in this paper we assume the best cold-start latency scenario when snapshots are stored on a local SSD.

%Agache et al. demonstrated that the cold-start latency is dominated by the host-based overheads as compared to the cluster infrastructure\cite{agache:firecracker}. 

For our analysis we launch function instances in our serverless framework, using the state-of-the-art Firecracker VM snapshotting feature~\cite{fc:snaps}.
(\marios{some methodology explanation})
For our latency experiments we collect times between XXX and XXX and run each function XXX times.
For our memory related experiments we collect data using...\marios{talk about ps and userspace faults here}
\end{comment}

%Tradeoff: latency vs memory utilization (cost-efficiency). Corroborating the prior work, we find that the dominant serverless latency part resides on the hosting server, as cluster infrastructure, namely the scale-out front-end and scale-up high-throughput low-latency worker manager, is able to process requests in under 30ms~\cite{agache:firecracker}. 

\subsection{Quantifying Cold-Start Delays}
\label{sec:char_lat}

We start by evaluating the cold-start latency of each function under study and compare it to the invocation latency of the warm function instance. Recall that a warm instance is memory-resident and does not experience any cold-start delay when invoked.
To obtain a detailed cold-start latency breakdown, we instrument the vHive-CRI orchestrator and invoke each function 10 times. To model a cold invocation, we flush the host OS page caches after each measurement.

%We compare cold and warm function invocations and try to attribute the performance difference.
%In a warm function invocation the function instance remains in memory after booting and the VM keeps running.
%For a cold invocation the infrastructure needs to allocate a new VM and boot the appropriate snapshot.
Figure~\ref{fig:lat_char} shows the latency for the cold and warm invocations for each function.
As expected, when a function instance remains warm (i.e., stays in memory), the instance delivers a very low invocation latency. By contrast, we find that a cold invocation from a snapshot takes one to two
%\marios{not consistent with the abstract that says 95\% overhead} 
orders of magnitude longer than a warm invocation, which indicates that even with state-of-the-art snapshotting, cold-start delays are a major pain point for functions.

To investigate the performance difference, we examine the end-to-end cold invocation latency breakdown. First, the vHive-CRI orchestrator spawns a new Firecracker process and restores the virtual machine monitor (VMM) state as well as the state of the emulated network and block devices -- the \emph{Load VMM} latency component. After that, the orchestrator resumes the loaded function instance's virtual CPUs and restores the persistent gRPC connection to the gRPC server inside the VM. We name this latency component as \emph{Connection restoration}. These two latency components are universal across all functions as they are part of the serverless infrastructure. Finally, we measure the actual function invocation processing time, referred to as \emph{Function processing}. 

The per-function latency breakdown is also plotted in Figure~\ref{fig:lat_char}. We observe that the first two universal components, namely \emph{Load VMM} and \emph{Connection restoration}, take \cor{156-317 ms}. Meanwhile, the actual function processing takes much longer (\cor{95\%} longer on average) for cold invocations as compared to warm invocations of the same functions, reaching into 100s of milliseconds even for functions like \texttt{helloworld} and \texttt{pyaes} that take only a few milliseconds to execute when warm.

The state-of-the-art snapshotting techniques rely on lazy paging to eliminate the population of guest memory from the critical path of VM restoration~(\S\ref{sec:cold-start101}). A consequence of this design is that each page touched by a function must be faulted in at the first access, resulting in thousands of page faults during a single invocation of a function. Page faults are processed serially because the faulting thread is halted until the OS brings the memory page from disk and installs it into the virtual address space by setting up the memory mappings in the process page table. In this case, the performance of the guest significantly depends on the disk latency as the OS needs to bring the missing pages from the guest memory file. %\boris{disk is true, but you also have the OS overhead of ioctl and installing page mappings.. why are you only mentioning the disk?}\dmitrii{disk/filesystem latencies are around 200usec while the rest are single/double-digit usec}

\begin{figure}[t]
\centering
\includegraphics[width=0.47\textwidth]{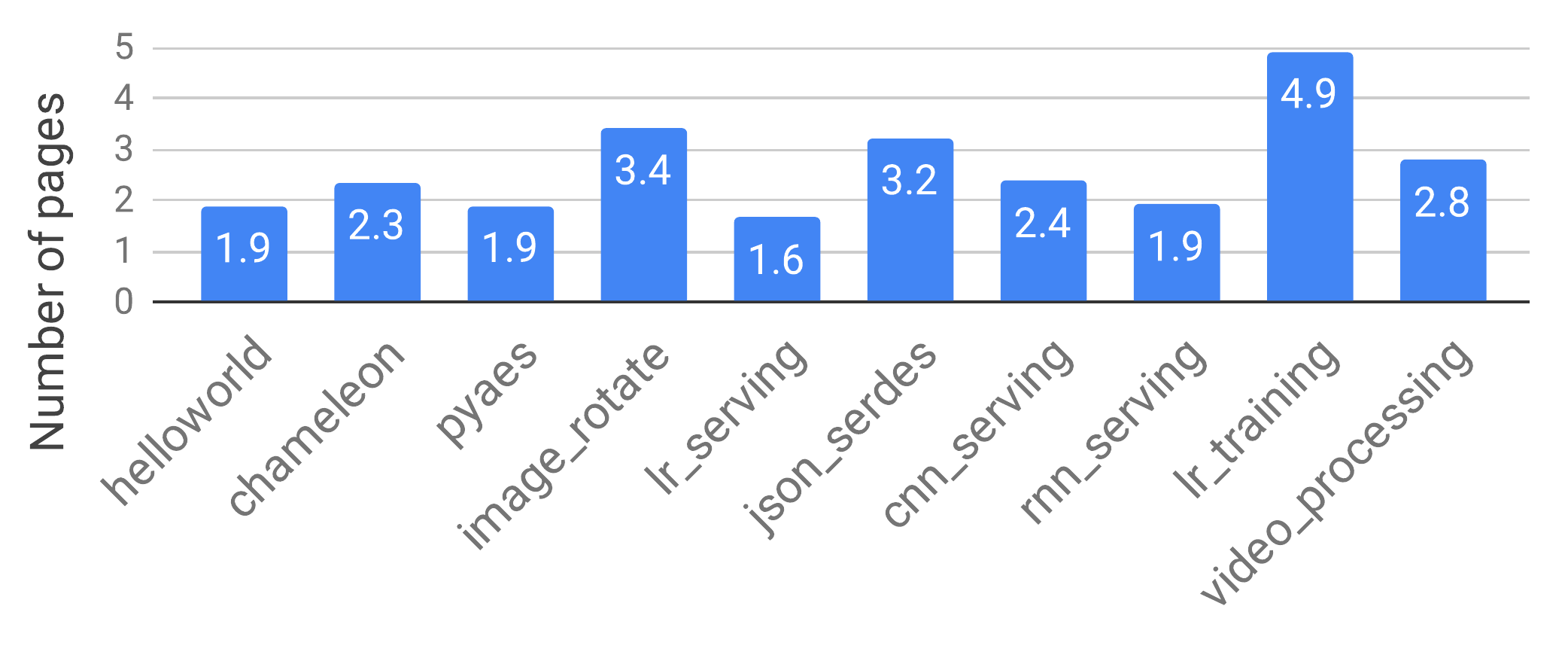}
\vspace{-15pt}
\caption{\cor{Guest memory pages contiguity.}}
\label{fig:char_contiguity}
\end{figure}

We also study the contiguity of the faulted pages, with the results depicted in Fig.~\ref{fig:char_contiguity}.
We find that function instances tend to access pages that are located in non-adjacent locations inside the guest memory. This lack of spatial locality significantly increases disk access time, and thus page fault delays, because sparse accesses to disk cannot benefit from the host OS's run-ahead prefetching. \cor{Fig.~\ref{fig:char_contiguity} shows that the average length of the contiguous regions of the guest-physical memory is around 2-3 pages for all functions except \texttt{lr\_training} that shows contiguity of up to 5 pages. }

%As a result, the pages are scattered across the guest-memory backing file inducing high page fault delays because sparse accesses to disk cannot benefit from the host OS's run-ahead prefetching.

%Firecracker snapshoting maps the guest memory as a file-backed memory region relying on lazy paging that, as we find, comes at a significant overhead due to thousands of page faults that arise during a single invocation of a function.

\begin{comment}
dive into the way Firecracker works.
Figure~\ref{fig:lat_snap} also shows the latency break down between the actual function processing, loading the VMM, and restoring the data path TCP connection from the gateway service to the function instance.
The last two are a fixed cost to be paid for every new VM allocation.
More importantly though, the actual function processing takes much longer in the cold case.
Firecracker snapshoting maps the guest memory as a file-backed memory region relying on lazy paging that, as we find, comes at a significant overhead due to thousands of page faults that arise during a single invocation of a function.
\end{comment}

%I assume the last sentence (restore TCP) is related to 'reconnect to function' in the graph. Make sure you use the same terminology in the graph and in the text. }

\subsection{Function Memory Footprint \& Working Set}
\label{sec:char_mem_footprint}

The above analysis demonstrates the benefits of keeping functions warm, because cold function invocations significantly increase the end-to-end function invocation latency.
In this subsection, we show that despite the advantages of warm functions, keeping many functions warm is wasteful in terms of memory capacity.
%\marios{i wouldn't say expensive. That's not your point. I would say wasteful because they are not necessary}.
%identify the potential memory pressure such a decision may create, showcasing the trade-off between cold-start delays and memory footprint. 

\begin{figure}[t]
\centering
\includegraphics[width=0.47\textwidth]{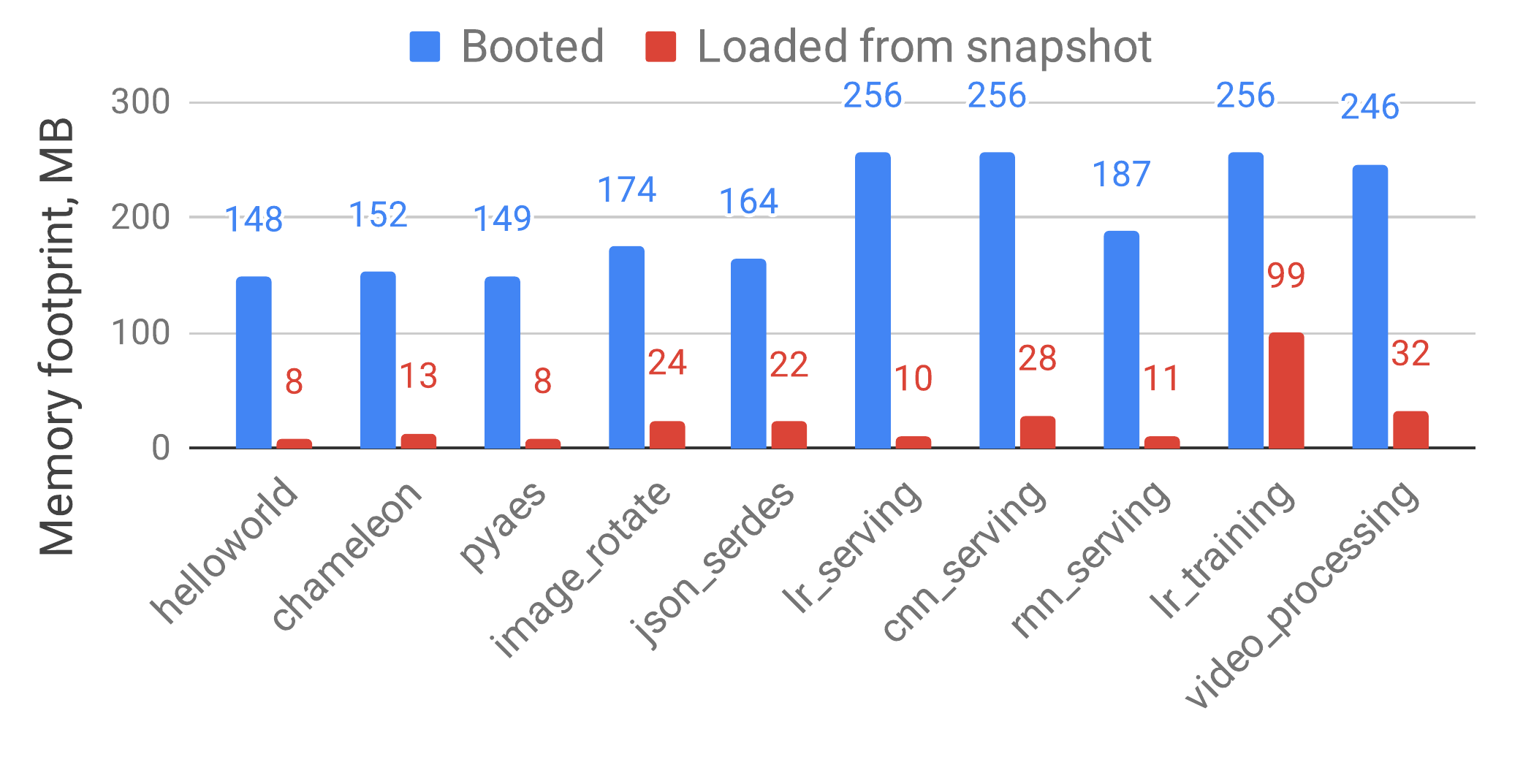}
\vspace{-15pt}
\caption{\cor{Memory footprint of function instances after one invocation.}}
\label{fig:char_footprint}
\end{figure}

%In fact, programming with serverless functions encourages developers to organize their services as collections of small and short-running software components.

%We also identify whether this memory footprint is related to the actual function invocation or the environment initialization. Specifically, we are interested in the number of pages a function touches when booted from a snapshot. Our intuition is that this number of pages will be significantly smaller, since a function instance engages much more logic in the guest OS and user-defined initialization part of the functions during the boot time as compared to processing the actual function invocation request.

We first investigate the fraction of a function instance's memory footprint that is related to the actual function invocation.
%To report the memory footprint of a VM, we use the Linux \texttt{ps} command, since a VM appears as a regular process for the host OS. We report the footprint as a number of distinct guest-physical memory pages that a function instance accesses during processing of an invocation request. 
\cor{First, we measure the total footprint of a booted VM after the first function invocation is complete using the Linux \texttt{ps} command, since a VM appears as a regular process in the host OS. This footprint includes the hypervisor and the emulated layer overhead (around 3MB~\cite{agache:firecracker}), the memory pages that are accessed during the VM's boot process, function initialization, and the actual invocation processing.
Second, for a VM that is loaded from a snapshot, we trace the pages, using Linux \texttt{userfaultfd}~\cite{man:userfaultfd}, that are accessed only during the invocation processing, i.e., from the moment the VM is loaded to the moment when the vHive-CRI orchestrator receives the response from the function. Unlike the first measurement, this footprint relates only to the invocation processing.
}

Figure~\ref{fig:char_footprint} (the blue bars) shows the memory footprint of a single freshly-booted function instance.  
We observe that, for all functions, their memory footprint reaches 100-200MB. 
Thus, assuming that a serverless provider co-locates thousands of different functions instances on the same host and disallows memory sharing for security reasons (as is the case in practice~\cite{agache:firecracker}), the aggregate footprint of function instances will reach into hundreds of gigabytes. 

Figure~\ref{fig:char_footprint} also plots the footprint of the function instances loaded from a snapshot after the first invocation (red bars).
%We observe that, in this case, the functions' working sets span 8-34MB -- a mere 4-13\% of their memory footprint after booting. 
\cor{We observe that, in this case, the functions' working sets span 8-99MB (24MB on average), which is 3-39\%, and 9\% on average of their memory footprint after booting. }
The reason why the memory footprint of a function booted from scratch is much higher than the one loaded from a snapshot is that starting an instance by booting requires many steps: booting a VM, starting up the Containerd's agents~\cite{fc-ctrd:deep-dive} as well as user-defined function initialization. This complex boot procedure engages much more logic (e.g., guest OS and userspace code) than just processing the actual function invocation, which naturally affects the former's memory footprint. 
%Based on these findings, we argue that a function's working set of guest memory pages is relatively compact even for a fully-virtualized function instances in a production-grade Containerd infrastructure. 

Despite the fact that, when loaded from a snapshot, the memory footprint of a function instance is relatively compact, the total memory footprint for thousands of such functions would still comprise tens of GBs. While potentially affordable memory-wise, we note that keeping this much state in memory is wasteful given the low invocation frequency of many functions (\S\ref{sec:serv_characteristics}). Moreover, such a high memory commitment would preclude colocating memory-intensive workloads on nodes running serverless jobs, thus limiting a cloud operator's ability to take advantage of idle resources. We thus conclude that while functions loaded from a snapshot present an opportunity in terms of their small memory footprint, by itself, they are not a solution to the memory problem.

\subsection{Guest Memory Pages Reuse}
\label{sec:mem_locality}

%\dmitrii{We trace page faults to the guest-physical memory, which is a mere virtual memory region of the Firecracker process, using Linux \texttt{userfaultfd} feature~\cite{man:userfaultfd} that allows a userland process to inspect the addresses and serve the page faults on behalf of the host OS.}

After establishing that the working sets of serverless functions booted from a snapshot are compact, we study how the working set of a given function changes across invocations. Our hypothesis is that the stateless nature of serverless functions results in a stable working set across invocations. 

User and guest kernel code pages account for a large fraction of functions' footprint.
This code belongs either to the underlying infrastructure or the actual function implementation.
Providers deploy additional control-plane services inside a function's sandbox and use general-purpose communication fabric (e.g., gRPC) to connect functions to the vHive-CRI orchestrator~\cite{agache:firecracker,fc-ctrd:deep-dive}. 
The gRPC framework uses the standard TCP network protocol, similarly to AWS Lambda~\cite{agache:firecracker}, that adds the guest OS's network stack to the instance footprint. 
Using the \texttt{helloworld} function, we estimate that this infrastructure overhead accounts for up to 8MB of a function's guest-memory footprint and is stable across function invocations.  

We observe that functions naturally use the same set of memory pages while processing different inputs. For example, when rotating different images or evaluating different customer reviews, functions use the same calls to the same libraries and rely on the same functionality inside the guest kernel, e.g., the networking stack. 
Moreover, the functions engage the same functionality that is a part of the provider's infrastructure, e.g., the Containerd's agents inside a VM~\cite{fc-ctrd:deep-dive}. 
\shep{Finally, we observe that even when a function's code performs a dynamic allocation, the guest OS buddy allocator is likely to make the same or similar allocation decisions.
These decisions are based on the state of its internal structures (i.e., lists of free memory regions), which is the same across invocations being loaded from the same VM snapshot. Hence, the lack of concurrency and non-determinism inside the user code of the functions that we study results in a similar guest physical memory layout.}
%ORIGINAL Finally, we observe that even when a function's code performs a dynamic allocation, the guest OS buddy allocator is likely to make the same or similar allocation decisions as its decisions are based on the state of its internal structures (i.e., lists of free memory regions), which is the same across invocations being loaded from the same VM snapshot.
\begin{figure}[t]
\centering
\includegraphics[width=0.47\textwidth]{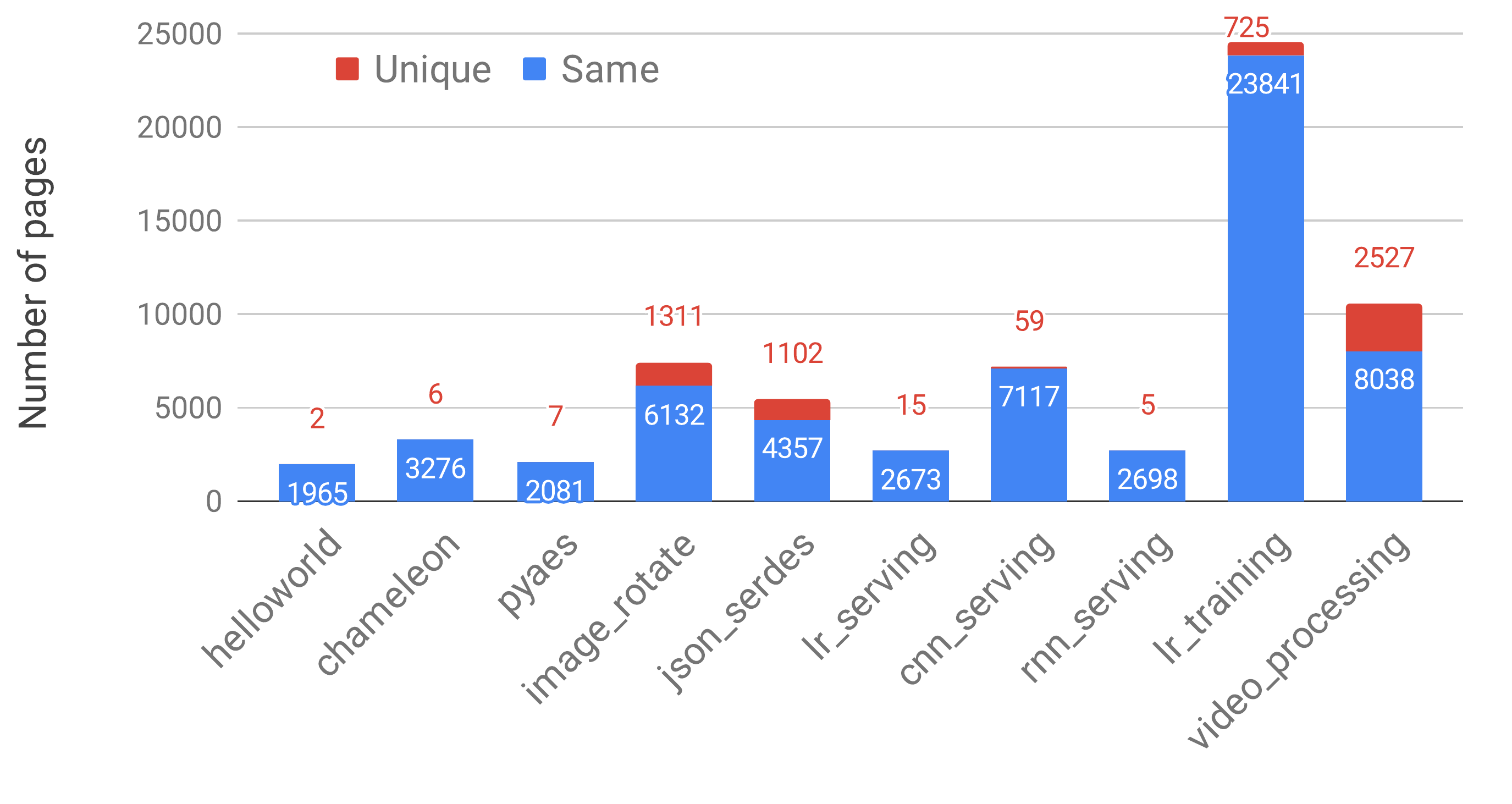}
\vspace{-15pt}
\caption{\cor{Number of pages that are unique or same across invocations with different inputs. The numbers above the bars correspond to uniquely accessed pages.}}
\label{fig:char_pages_reuse}
\end{figure}

%We validate our hypothesis\marios{you kept saying observe above. So, it was not a hypothesis, but the result of experimentation. Either fix the observe-s, or remove the hypothesis, or move this para earlier.} \boris{This is a tricky one. The section does start out by clearing stating a hypothesis, and the observations help explain the basis for it. But by the time the reader gets to this point in the text, them may have forgotten about the hypothesis that appeared several paragraphs earlier. Moving this paragraph earlier would be the best, if it's easy enough without disrupting the flow and keeping the obseervations somewhere nearby. Overall, I'd say this is not a major item imo.} 
We validate our hypothesis about the working sets by studying the guest memory pages that are accessed when a function is invoked with different inputs. 
%speculation by that functions exhibit strong reuse patterns in accessing guest memory \textit{across} function invocations by \marios{explain the experiment! How many invocations? How input was different etc}
Fig.~\ref{fig:char_pages_reuse} demonstrates that the majority of pages accessed by all studied functions are the same across invocations with different inputs. 
%For 7 out of 9 functions, more than 97\% of the memory pages are the same across invocations.
\cor{For 7 out of 10 functions, more than 97\% of the memory pages are identical across invocations.
For \texttt{image\_rotate}, \texttt{json\_serdes}, \texttt{lr\_training}, and \texttt{video\_processing}, reuse is lower because these functions have large inputs (photos, JSON files, training datasets, and videos, respectively) that are 1-10MB in size. Nonetheless, even for these three functions, over 76\% of memory pages are the same across invocations.}
% image-rotate and lr-training:  76\% and 72\% correspondingly

%\boris{Why disk? I thought a function is either in memory or is loaded over a network if cold booted. I didn't see snapshots introduced (may be I missed it).}

\subsection{Summary}

%\marios{How about turning this to list of conclusions, rather than text?}
We have shown that invocation latencies of cold functions may exceed those of warm functions by one to two orders of magnitude, even when using state-of-the-art VM snapshots for rapid restoration of cold functions. We found that the primary reason for these elevated latencies is that the existing snapshotting mechanisms populate the guest memory on-demand, thus incurring thousands of page faults during function invocation. These page faults are served one-by-one by reading non-contiguous pages from a snapshot file on disk. The resulting disk accesses have little contiguity and induce significant delays in processing of these page faults, thus slowing down VM restoration from a snapshot.

We have further shown that function instances restored from a snapshot have compact working sets of guest memory pages, spanning just \cor{24MB, on average}. Moreover, these working sets are stable across different invocations of the same function; indeed, the function instances access predominantly the same memory pages even when invoked with different inputs. 

%page faults to the guest-physical memory backing file dominate the cold-start latency of the serverless functions that we studied. These page faults trigger a large number of time-consuming sparse disk accesses to the guest memory file. However, our main insight is that functions use a small number of memory pages during their invocation and they reuse a large fraction of these pages across invocations. That provides an opportunity for prefetching guest memory pages from the disk into main memory before the VM requests them directly and it is the main idea behind \tac{}.

%\input{4-design}
\section{\tac{}: \tnameabs{}}
\label{sec:design}

The compact and stable working set of a function's guest memory pages, which instances of the given function access across invocations of the function, provides an excellent opportunity to slash cold-start delays by prefetching. 

\cor{Based on this insight, we introduce {\tname}~({\tac}), a light-weight software mechanism inside the vHive-CRI orchestrator to accelerate
%Based on this insight, we introduce {\tname}~({\tac}) -- a vHive orchestrator\marios{why vHive orchestrator?} \boris{how about "REAP, a light-weight software mechanism to accelerate..."} that accelerates 
function invocation times in serverless infrastructures. 
}
REAP records a function's working set upon the first invocation of a function from a snapshot and replays the record to accelerate load times of subsequent cold invocations of the function by eliminating the majority of guest memory page faults. The rest of the section details the design of \tac.

\subsection{{\tac} Design Overview}
\label{sec:design_overview}

\begin{figure}[t]
\centering
\subfloat[{\tac} record phase.]{
    \includegraphics[width=0.40\textwidth]{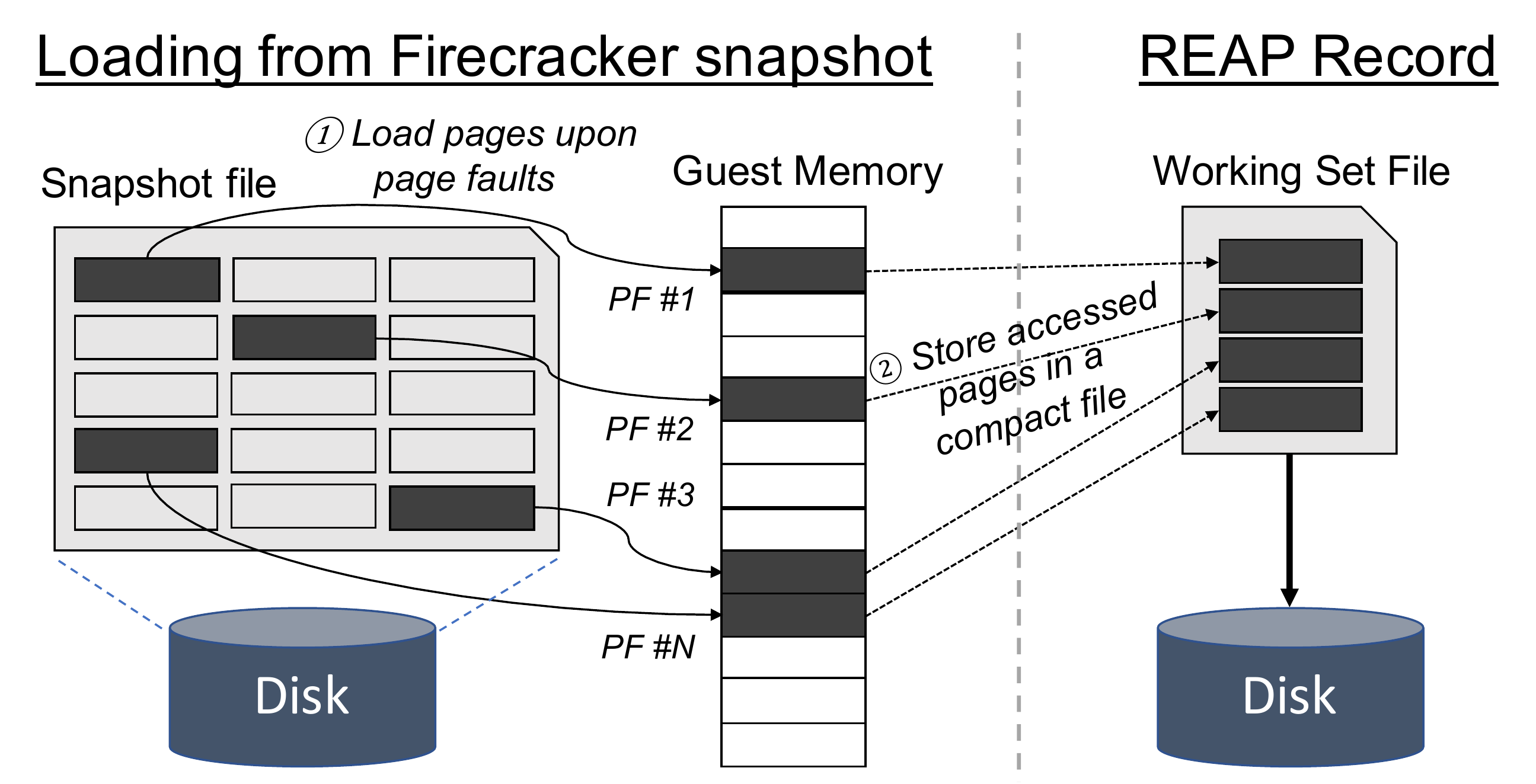}
    \label{fig:design_record}}
\newline
\centering

\subfloat[{\tac} prefetch phase.]{
    \includegraphics[width=0.40\textwidth,trim={2cm 4.5cm 2.5cm 5cm},clip]{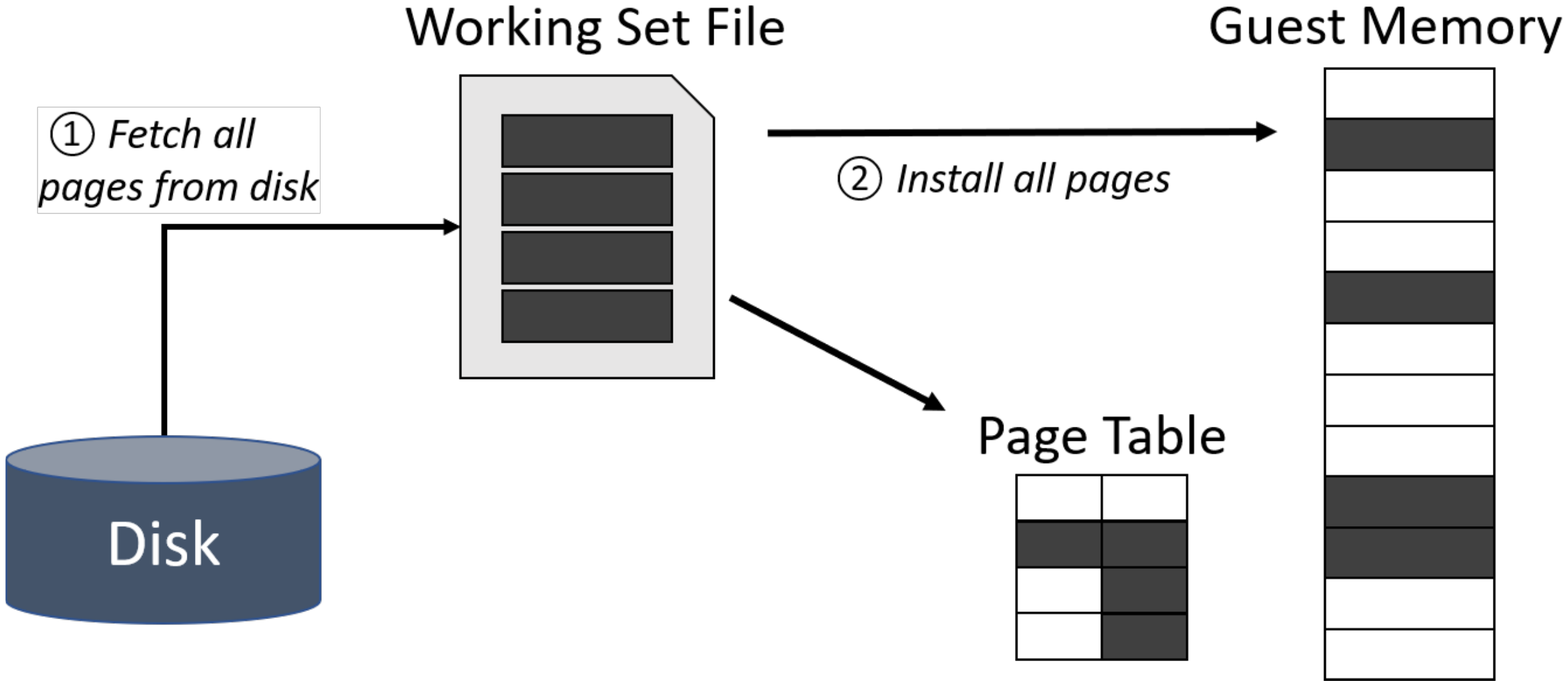}
    \label{fig:design_prefetch}}
\caption{\tac's two-phase operation.}
\label{fig:design_overview}
\end{figure}

\begin{comment}

\begin{figure}[t]
\centering
\includegraphics[width=0.47\textwidth,trim={3cm 6cm 5cm 4cm},clip]{figs/design_record.pdf}
\caption{{\tac} record phase.}
\label{fig:design_record}
\end{figure}

\begin{figure}[t]
\centering
\includegraphics[width=0.47\textwidth,trim={6cm 5.5cm 8cm 5.5cm},clip]{figs/design_prefetch.pdf}
\caption{{\tac} prefetch phase.}
\label{fig:design_prefetch}
\end{figure}

\end{comment}

Given an existing function snapshot, {\tac} operates in two phases. During the \emph{record} phase, {\tac} traces and inspects the page faults that a function instance triggers when accessing pages in the guest memory, identifying the positions of these pages in the backing guest memory file (Fig.~\ref{fig:design_record}). After a function invocation is complete, \tac{} creates two files, namely the \emph{working set (WS) file)} that contains a copy of all accessed guest memory pages in a contiguous compact chunk and the \emph{trace file} that contains the offsets of the original pages inside the guest memory file.
%file with  which we refer to as the \emph{working set (WS) file}, that contains a copy of all accessed guest memory pages in a contiguous compact trace.
The contiguous compact WS file can be rapidly brought into physical memory in a single read operation, which greatly reduces disk and system-level overheads in the snapshot baseline that requires many disparate accesses to pages scattered across the guest memory file on disk.

After the completion of the record phase, all future invocations of the function enjoy accelerated load times as \tac's {\em prefetch} phase eagerly populates the guest memory from the WS file before launching the function instance 
%In the \emph{prefetch} phase that comes after the record for a function is available, {\tac} accelerates all future cold invocations of that function by populating the guest memory right before loading a new instance of that function 
(Fig.~\ref{fig:design_prefetch}). Upon an arrival of a new invocation, {\tac} fetches the entire WS file from disk into a temporary buffer in the orchestrator's memory and eagerly installs the pages into the function instance's guest memory region. \tac{} also populates the page table of the instance in the host OS. As a result, when the instance is loaded, the function executes without triggering page faults to the stable memory working set. Page faults to uniquely-accessed pages in a given invocation are handled by REAP on demand.

\subsection{Implementation}
\label{sec:implementation}

\tac{} adheres to the following design principles, which facilitate adoption and deployment in a cloud setting: i) REAP is agnostic to user codebase; ii) \tac{} is independent of the underlying serverless infrastructure; iii) \tac{} is implemented entirely in userspace without kernel modifications; iv) \tac{} works efficiently in a highly multi-tenant serverless environment.

\vspace{3pt}
We implement REAP as a part of the vHive-CRI orchestrator that controls the lifecycle of all function instances. For each function, the vHive-CRI orchestrator tracks active function instances and performs the necessary bookkeeping, including maintaining the snapshot files and working set records. To accommodate the highly-concurrent serverless environment with many function instances executing simultaneously, it is a fully parallel implementation with a dedicated \emph{monitor} thread for each function instance. 
Each monitor thread records or prefetches the working set pages and also serves page faults that are raised by the corresponding instance. In our prototype, the monitor threads are implemented as lightweight goroutines, which are scheduled by the Go runtime. 

To implement the monitor, we use the Linux \texttt{userfaultfd} feature that allows a userspace program to handle page faults on behalf of the OS. In Linux, a target process can register a virtual memory region in anonymous memory and request a user-fault file descriptor, which can be passed to a monitor running as a separate thread or process. The monitor polls for page fault events that the OS forwards to the user-fault file descriptor. Upon a page fault, the monitor installs the contents of the page that triggered the page fault. The monitor is free to retrieve page contents from any appropriate source, such as a file located on a local disk or from the network. Furthermore, the monitor can install any number of pages before waking up the target process. %\boris{Please check:} 
Thanks to these features, the monitor can support both local and remote snapshot storage, and can eagerly install the content of the entire WS file at once. 

Upon each function invocation for which there is no warm instance available, the vHive-CRI orchestrator launches the monitor thread in one of two modes: record, if no WS file is available for this instance, or prefetch, if a corresponding WS file exists. 

\begin{comment}
\subsubsection{Monitor Process}

Every active function instance is associated with a monitor process.
The monitor behaves differently depending on whether it is the first invocation of the function or a subsequent one and enters the record or the prefetch mode accordingly.
The monitor runs on the hypervisor.

% userlevel page faults
The main mechanism behind \tac{} is userlevel page fault handling through \texttt{userfaultfd}~\cite{man:userfaultfd}.
Every time  during the function execution there is a page fault, the monitor is notified through an event file descriptor and acts accordingly depending on whether it is the record or the prefetch phase.

\end{comment}

\subsubsection{Record phase}
The goal of the monitor during the record phase is to capture the memory working set for functions instantiated from snapshots.
%A VM snapshot contains a file that stores the contents of the entire guest memory.
Before loading the VMM state from a snapshot, the hypervisor maps the guest memory file as an anonymous virtual memory region and requests a user-fault file descriptor from the host OS, passing this descriptor over a socket to the monitor thread of the vHive-CRI orchestrator. Then, the hypervisor restores the VMM and emulated devices' state and resume the virtual CPUs of the newly loaded function instance that can start processing the function invocation.

Every first access to a guest memory page raises a page fault that needs to be handled and recorded by the monitor.
The monitor maps the guest memory file as a regular virtual memory region in the monitor's virtual address space and polls (using the \texttt{epoll} system call) for the host kernel to forward the page fault events, triggered by the instance. 
Upon receiving a page fault event, the monitor reads a control message from the user-fault file descriptor that contains the description of the page fault, including the address in the virtual address space of the target function instance. The monitor translates this virtual address into an offset that corresponds to the page location in the guest memory file. While serving the page faults, the monitor records the offsets of the working set pages in the trace file.

We augment the Firecracker hypervisor to inject the first page fault of each instance to the first byte of the instance's guest memory. Doing so allows file offsets for all of the following page faults to be derived by subtracting the virtual address of the first page fault. 
Using the file offset of the missing page, the monitor locates the page in the guest memory file and installs the page into the guest memory region of the instance by issuing an \texttt{ioctl} system call to the host kernel, which also updates the extended page tables of the target function instance. 
After the vHive-CRI orchestrator receives a response from the function, indicating that the function invocation processing has completed, the monitor copies the recorded working set pages, using the offsets recorded in the trace file,
%\boris{I didn't see it mentioned above that something is being recorded.} 
into a separate WS file (\S\ref{sec:char_mem_footprint}).
%Together the recorded pages the working set of pages, inside the guest-memory file, that a function instance needs to process an invocation.

%Figure~\ref{fig:record} summarizes the record phase and the interaction between the function execution and the monitor. \marios{explain fig}.

Note that the record phase increases the function invocation time as compared to the baseline snapshots due to userspace page fault handling. As such, \tac{} penalizes the first function invocation to benefit subsequent invocations. We quantify the recording overheads in \S\ref{sec:eval_record}.

\subsubsection{Prefetch phase}
For every subsequent function invocation, the vHive-CRI orchestrator spawns a dedicated monitor thread that
%\boris{This sentence didn't make sense, so I rewrote. Please check.}\sout{uses the recorded pages to prefetch the working set memory pages by reading the guest memory file, located on disk, into a buffer in the monitor's memory with a single \texttt{read} system call.}
uses the WS file to prefetch the working set memory pages from disk into a buffer in the monitor's memory with a single \texttt{read} system call.
Then, the monitor eagerly and proactively installs the pages into the guest memory through a sequence of \texttt{ioctl} calls, following which it wakes up the target function instance with another \texttt{ioctl} call.
%This prefetch phase is overlapped with loading the VMM to further improve start-up latency.

As in the record phase, the monitor maps the guest memory file during every subsequent cold function invocation. After installing all the working set pages from the WS file, the monitor keeps polling for page faults to pages that are missing from the stable working set and installs them on demand, as in the record phase. Since the WS file captures the majority of pages that a function instance accesses during an invocation, only a small number of page faults needs to be served by the monitor on demand.
%\marios{not consistent with the end of 5.1 that said they are handled the same as without REAP}.
%it keeps running even after the prefetch phase throughout the function execution serving page faults for pages that do not belong in the working set, thus not proactively installed.

%Figure~\ref{fig:design_prefetch}...\marios{describe the fig}.

\subsubsection{Disk Bandwidth Considerations}
\label{sec:design-final}

\tac{}'s efficiency depends entirely on the performance of the prefetch phase and, specifically, how fast the vHive-CRI orchestrator can retrieve the working set pages from disk. Although a single commodity SSD can deliver up to 1-3 GB/s of read bandwidth, SSD throughput varies considerably depending on disk access patterns. 
An SSD can deliver high bandwidth with one large multi-megabyte read request, or with $>$10 4KB requests issued concurrently.
For example, on our platform (\S\ref{sec:method}), with a standard Linux \texttt{fio} IO benchmark~\cite{fio} that issues a single 4KB read request, the SSD can deliver only 32MB/s, whereas issuing 16 4KB requests can increase the SSD throughput to 360MB/s. While concurrent reads deliver much higher bandwidth than a single 4KB read, the achieved bandwidth is still considerably below the peak of 850MB/s of our Intel SATA3 SSD. 

We find that REAP achieves close to the peak SSD read throughput (533-850MB/s) by fetching the WS file in a single $>$8MB read operation that bypasses the host OS' page cache (i.e., the WS file needs to be opened with the \texttt{O\_DIRECT} flag). 
%With these performance considerations in mind, we showcase the working set retrieval optimizations that allow REAP to leverage full SSD bandwidth in \S\ref{sec:eval_opts}.

\subsection{Discussion}
\label{sec:discussion}

\tac{} adheres to the design principles set out in \S\ref{sec:implementation}. We implement REAP entirely in userspace as a part of the vHive-CRI orchestrator.
It is written in 4.5K Golang LoC, including tests and benchmarks, and is loosely integrated with the industry-standard Containerd framework~\cite{fc-ctrd:github,containerd:industry,fc-ctrd:deep-dive} via gRPC.
%REAP can be seamlessly integrated with any infrastructure that is compliant with Containerd (e.g., Kata Containerd~\cite{kata:containers}, gVisor~\cite{google:gvisor,ctrd:gvisor}, Kubernetes~\cite{k8s})\marios{remove the sentence? This should be only about REAP}.
The implementation does not require any changes to host or guest OS kernel. 
We add less than 200 LoC to Firecracker's Rust codebase,
%\boris{Please factor out the rest of this sentence into a stand-alone sentence:} 
not including two publicly available Rust crates that we used, to register a Firecracker VM's guest memory with \texttt{userfaultfd} and to delegate page faults handling to the vHive-CRI orchestrator. Finally, the orchestrator follows a purely parallel implementation by spanning a lightweight monitor thread (goroutine) per function instance.

\section{Methodology}
\label{sec:method}

\subsection{Serverless Host Infrastructure}

Our host infrastructure is based on de facto industrial standard Containerd framework~\cite{containerd:industry,cncf,kata:containers,fc-ctrd:deep-dive,ctrd:gvisor} that we extend to support the newly introduced Firecracker snapshots functionality~\cite{fc:snaps}.
%We choose Containerd because of its modularity, as nearly all containerd components are independent plugins, and wide support of the open-source and industrial communities led by Cloud Native Computing Foundation ~\cite{cncf,kata:containers,fc-ctrd:deep-dive,ctrd:gvisor}.
We introduce a custom-build gateway service, which is similar to Amazon Lambda infrastructure~\cite{agache:firecracker}, that interacts both with control plane, i.e., Containerd, that manages the lifecycle of all instances, starting or loading them from a snapshot on demand, and acts as a software router that routes the function invocation requests to the appropriate instances and returns their responses. 
The gateway service establishes a persistent connection with Containerd for control messages and persistent connections with each active function instance. We choose gRPC fabric~\cite{google:grpc} as a common choice for a general-purpose communication framework for communicating to Containerd and all function instances.

We use a collection of benchmarks that drive our experiments that model the end-to-end cold-start latency. The latency includes all communication and processing steps, starting from issuing invocation requests to the gateway service that loads the appropriate instances on demand and forwards the requests to the function instances. Finally, the instances send the responses through all the aforementioned components back to the benchmark.

A function instance comprises a Firecracker microVM that that runs a Containerd internal control-plane agent and a gRPC server, running on top of Linux Alpine. Upon receiving an invocation, the gRPC server invokes a user-provided function upon an invocation and sends the function's response back to the gateway service.

\subsection{Evaluation Platform}

We conduct our experiments on a 2$\times$24-core Intel Xeon E5-2680 v3 , 256GB RAM, Intel 200GB SATA3 SSD, running Ubuntu 18.04 Linux with kernel 4.15. We set the CPU frequency to 2.5GHz to enable predictable and reproducible latency measurements. We disallow memory sharing among all function instances and disable swapping to disk, as suggested by AWS Lambda production guidelines~\cite{agache:firecracker,aws:prod}. 

We use a collection of Python-based functions (Table~\ref{tbl:functionbench}) that include a simple \texttt{helloworld} function and eight functions from the representative FunctionBench~\cite{kim:functionbench,kim:practical}, covering a wide range of application domains and frameworks.
The root filesystems for all functions are generated automatically by Containerd using devmapper-based overlay filesystem from Linux Alpine OCI/Docker images. We package function inputs (files, images, machine learning models, etc.) inside the OCI images, and use different inputs in the record and prefetch phases of {\tac}.

We optimize virtual machines for minimum cold-start delays, similar to production serverless setups~\cite{kaffes:centralized,agache:firecracker}. The VMs run a guest OS kernel 4.14 without modules. Each VM instance has a single vCPU. We boot VM instances with 256MB guest memory that is the minimum amount to boot all the functions 
%\boris{minimum for the bulk of the functions doesn't sound sufficient for all functions. Please clarify.}
that we adopt from FunctionBench.

\begin{table}[!t]
\vspace{3mm}
  \caption{Serverless functions adopted from FunctionBench.}
%\vspace{-4pt}
  \centering
  \small
%\begin{tabular}{| >{\raggedright\arraybackslash}m{12mm}| m{63mm}|}
\begin{tabular}{|l|l|}
    \hline
     Name    & Description \\
    \hline
    \hline
    helloworld & Minimal function\\
    \hline
    chameleon & HTML table rendering\\
    \hline
    pyaes & Text encryption with an AES block-cipher \\
    \hline
    image\_rotate & JPEG image rotation \\
    \hline
    json\_serdes & JSON serialization and de-serialization \\
    \hline
    lr\_serving &  Review analysis, serving (logistic regr., Scikit) \\
    \hline
    cnn\_serving & Image classification (CNN, TensorFlow) \\
    \hline
    rnn\_serving & Names sequence generation (RNN, PyTorch) \\
    \hline
    lr\_training & Review analysis, training (logistic regr., Scikit)\\
    \hline
    video\_processing & \cor{Applies gray-scale effect (OpenCV)}\\
    \hline
  \end{tabular}
%  \vspace{-7pt}
  \label{tbl:functionbench}
%  \vspace{-4pt}
\end{table}

We measure the cold-start latency of function instances from the moment when the invocation request reaches the gateway service
%, which forwards the request to the appropriate function instance, 
to the moment when the gateway service receives the response from the function instance. The latency includes both the control-plane delays (including interactions with Containerd and Firecracker hypervisor) and data-plane processing time that is gRPC requests processing and actual function execution. 
To simulate the low invocation frequency of serverless functions in production~\cite{shahrad:serverless}, we flush the host OS' page cache before each invocation of a cold function.
We instrument the gateway service to break down the measured end-to-end latencies for various configurations that we evaluate.

%, which interacts with Containerd-based control plane and acts as a software router for sending requests to and from function instances,

%\dmitrii{highlight that the baseline Firecracker snaphots is equivalent to the cold-start Catalyzer config due to lazy paging. We don't consider other Catalyzer option because memory sharing is commonly disallowed in production environment}

%\section{Evaluation}
%\label{sec:eval}

\subsection{Understanding REAP Optimizations}
\label{sec:eval_opts}

%\boris{The intermediate REAP configurations (bars 2 and 3) are incomprehensible in the text. In fact, the entire setup is hard to understand because the Design section makes no mention of optimization steps -- it just presents a single design. You need to (1) get rid of this notion of optimization steps, and (2) present bars 2 and 3 as itermediate design points and explain them much better. See my comments below.}

We start by evaluating the cold-start latency of the \texttt{helloworld} function, whose short user-level execution time is useful for understanding serverless framework overheads. 
In addition to evaluating the baseline Firecracker snapshots and \tac{} as presented in \S\ref{sec:design}, we also study two additional design points that help justify \tac's design decisions. Specifically, we consider the following configurations.

%Fig.~\ref{fig:eval_opt_steps} shows the cold-start latencies that are observed with different \tac{} configurations, including the baseline Firecracker snapshots and three \tac{} optimization steps. 

%and the configurations that follow the optimization steps described in \S\ref{sec:implementation}. 

\begin{figure}[t]
\centering
\includegraphics[width=0.47\textwidth]{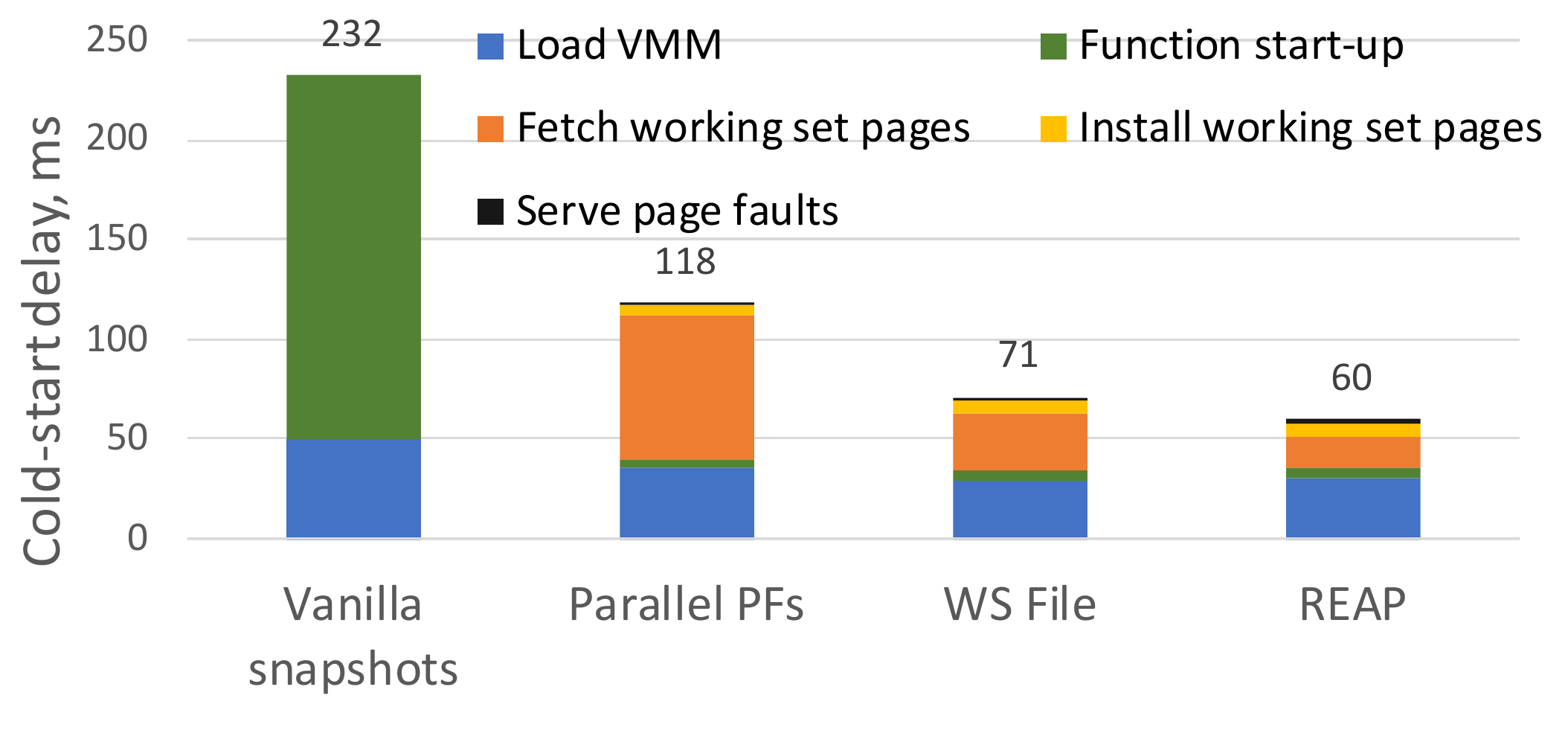}
\caption{REAP optimization steps.}
\label{fig:eval_opt_steps}
\end{figure}

{\em Vanilla snapshots:} This is the baseline configuration, which restores the VMM and the emulation layer in 50ms, then spends 182ms processing the function invocation (Fig.~\ref{fig:eval_opt_steps}) that takes just 1ms for for a warm instance (Fig.~\ref{fig:lat_char}).
%\boris{I'm confused. Why do you say that it's waiting for invocation? Fig 1 shows that it's a combination of connection restoration AND function processing. You need to be super consistent. Also, you might want to explicitly draw attention to Fig 1 and point out that warm processing time is 1ms.}
The large processing delay is directly attributed to the handling of page faults in the critical path of function execution.
%arises from the fact that the function instance's virtual CPU thread accesses one page at a time, which significantly slows down function invocation because the host OS has to serve all page faults one at a time. 
As \texttt{helloworld}'s working set is around 8MB, one can infer that vanilla snapshotting is only able to utilize 43MB/s of SSD bandwidth, i.e., $<$5\% of the peak bandwidth on our platform.

{\em Parallel Page Fault handling:} This design (labeled ``Parallel PFs'' in Fig.~\ref{fig:eval_opt_steps}) parallelizes page fault processing. It does so by using  the trace file specifying the offsets of the pages comprising the stable working set, and deploys goroutines to bring in the associated pages, in parallel, from the guest memory file. 
For this and all the following configurations, we make 16 hardware threads available to vHive-CRI goroutines, managed by the Go runtime.
%\marios{is this different that the other configs?}. 
Note that this configuration does {\em not} use the WS file. 

We observe that parallelizing page faults reduces function invocation time by 1.9$\times$ (to 118ms) by overlapping I/O processing and exploiting SSD's internal parallelism.
Repeating the same calculation as for the baseline, we identify that the orchestrator uses only 130MB/s of SSD bandwidth, which is 15\% of the maximum.  
This design point underlines that achieving high read bandwidth from the SSD is key to efficient page fault processing, and that lowering software overheads by itself is insufficient.

{\em WS file:} This design leverages the WS file (Sec~\ref{sec:design_overview}), which enables fetching the entire stable memory working set of a function with a single IO read operation. The difference between this design point and REAP is that the former reads into the OS page cache (which is the default behavior in Linux), whereas REAP bypasses the page cache (\S\ref{sec:design-final}).
%design point that seeks to optimize the SSD bandwidth bottleneck, replacing many parallel page-sized disk, as in \emph{Fetch WS}, accesses with a single read IO operation. Hence, similarly to the final \tac{} design, this design point fetches the working set pages from the separate \emph{condensed} working set file rather than the guest memory file, as in \emph{Fetch WS} design point.
From the figure, one can see that fetching the pages from the WS file can be performed in 29ms, 3.1$\times$ faster than through parallel page-sized reads (``Parallel PFs'' bar in Fig~\ref{fig:eval_opt_steps}). This design point utilizes 275MB/s of SSD bandwidth. 
%%We observe that serving the page faults to the pages that are not a part of the function working set, and hence missing from the WS file, takes more time, as compared to ``Parallel PFs''.
%\marios{not very clear paragraph}We observed that retrieving and installing the pages required by the function and that are not part of the WS file (2 pages, on average, per function instance as shown in Fig.~\ref{fig:char_pages_reuse}) takes more time than for the previous\marios{what does previous refer to?} configuration. \boris{Previous sentence is impossible to make sense of.} 
%%We find that the first few page faults, which happen after a period of low disk utilization, are more likely to become stragglers (specifically, one out of the first few page faults can take up to 16ms in our system). After ruling out low-power modes as a likely culprit, we attribute this behavior to an inefficiency in the filesystem. This effect is not visible in ``Parallel PFs'' likely because that design point raises a multitude of concurrent page faults, masking the first straggler page faults.

{\em \tac{}:}
The last bar shows the performance of the actual REAP design, as described in Sec~\ref{sec:design-final}, that fetches the working set pages from the WS file and bypasses the OS page cache. As the figure shows, retrieving the working set pages is accelerated by 2$\times$ (to 15ms) over the ``WS File'' design point that does not bypass the page cache. This highlights that while it's essential to optimize for disk bandwidth, software overheads also cannot be ignored.
In this final configuration, \tac{} achieves 533MB/s of SSD bandwidth, which is within 37\% of the 850MB/s peak of our SSD.

%%%%%%%%%%%%%%%%%%%%%%%%%%%%%%%%%%%%%%%%%%%%%%%%%%%%%%%%%%%%%%
%% COMMENT
%%%%%%%%%%%%%%%%%%%%%%%%%%%%%%%%%%%%%%%%%%%%%%%%%%%%%%%%%%%%%%
\begin{comment}
, as the \emph{ERLI} configuration, but also installs all working set pages into the guest-physical memory virtual memory region upon the first page fault, eliminating $>$99\% of page faults during function invocation. This eager installation of the working set effectively eliminates the polling overhead of the previously discussed configurations, reducing the overall latency down to XXms. 

\plamen{Removed coalesce}

The next bar to the right, \emph{coalesce}, adds coalescing of the guest memory pages from the working set into a smaller number of contiguous regions. Installing the working set as regions requires fewer \texttt{IOCTL}-s, as compared to the page-by-page installation in the previous configurations, decreasing the total cold-start latency to XXms.

The final optimization is \emph{condensing} the working set pages by copying them from their scattered locations in the guest memory file. Fetching the working set of pages from a small contiguous file takes advantage of all available SSD bandwidth, cutting the prefetching overhead compared to \emph{ERLI} by 9.43$\times$ to 13ms, and the overall cold-start delay to XXms. 

\end{comment}
%%%%%%%%%%%%%%%%%%%%%%%%%%%%%%%%%%%%%%%%%%%%%%%%%%%%%%%%%%%%%%
%%%%%%%%%%%%%%%%%%%%%%%%%%%%%%%%%%%%%%%%%%%%%%%%%%%%%%%%%%%%%%

\subsection{\tac{} on FunctionBench}
\label{sec:eval_reap}

\begin{figure*}[t]
\centering
\includegraphics[width=\textwidth]{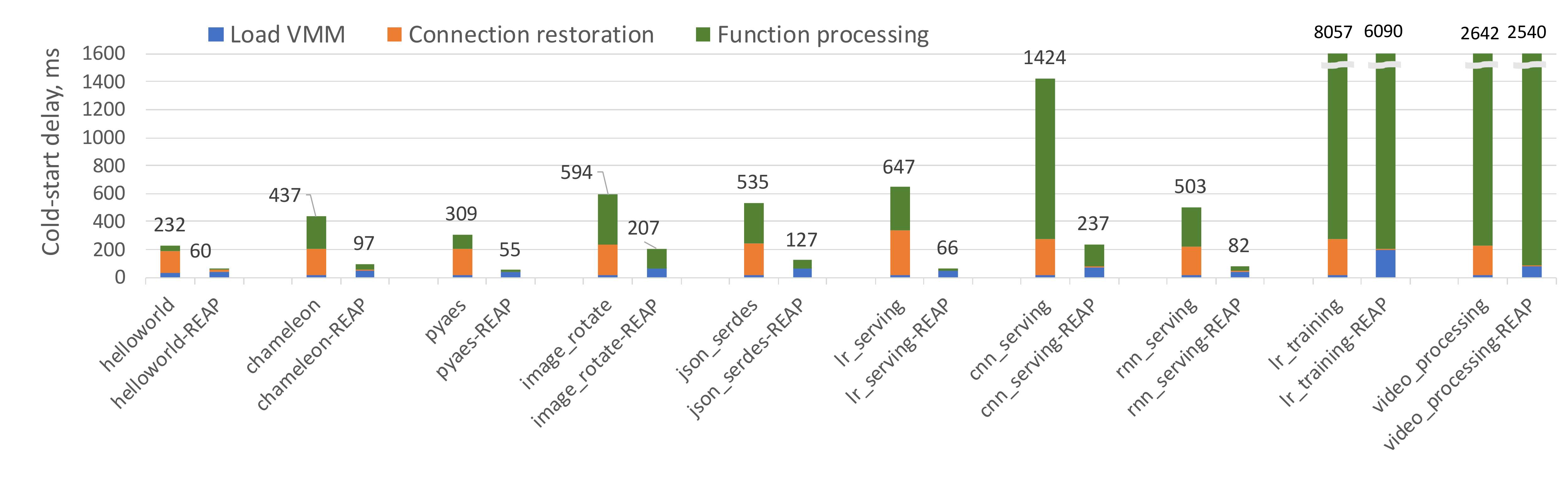}
\vspace{-25pt}
\caption{\cor{Cold-start delay with baseline snapshots (left bars) and REAP (right bars).} 
%\boris{This figure doesn't look good for us because LR  (last bars) are really tall. You should cut off the Y-axis at around 1000, and just label the LR bars.}
%\dmitrii{Added recording bar as well} \boris{workload names are messed up/missing. And where is recording?}
}
\label{fig:eval_one_instance}
\end{figure*}

%\boris{Fig 8 should be on this page; currently it's on the next one.}\dmitrii{sorry, not possible}

Fig.~\ref{fig:eval_one_instance} compares the cold-start delays of all the functions that we study with the baseline Firecracker snapshots and REAP prefetching. 
%With REAP, all functions' invocations become 1.1-4.5$\times$ faster (2.7$\times$ on average). 
\cor{With REAP, all functions' invocations become 1.04-9.7$\times$ faster (3.7$\times$ on average).}
%The fraction of time that accounts for restoring the connection from the orchestrator to the function's gRPC server shrinks by 20-30$\times$ to a mere 6-7ms thanks to existence of a stable working sets that accounts for this core functionality and which is prefetched by \tac{}.
%The function processing latency is reduced by 3.3$\times$, on average, for all functions except \texttt{image\_rotate}. In \texttt{image\_rotate}, function processing latency increases due to a large number of pages that contain the image input to the function and that vary from invocation to invocation (Fig.~\ref{fig:char_pages_reuse}). The orchestrator has to serve these missing pages one-by-one as page faults arise.\cor{RevA: improve the explanation}
The fraction of time for restoring the connection from the orchestrator to the function's gRPC server shrinks by \cor{45$\times$}, on average to a mere \cor{4-7ms} thanks to the stable working set for this core functionality that is prefetched by \tac{}.

\cor{Although we find that {\tac} efficiently accelerates the actual function processing, functions with a large number of pages missing from the recorded working set benefit less from {\tac}.
The function processing latency is reduced by 4.6$\times$, on average, for all functions except \texttt{video\_processing}. \shep{During the REAP record phase, the \texttt{video\_processing} function takes a video fragment of a different aspect ratio than in the prefetch phase that, as we suspect, changes the way OpenCV performs dynamic memory allocation (e.g., uses buffers of different sizes), resulting in a different guest physical memory layout and, hence, different working sets.}
%\shep{different dynamic allocation decisions of the buddy allocator when the function processes video fragments of different aspect ratios that we use in the REAP record and prefetch phases.}  \boris{I don't understand what the previous sentence is saying. That aspect ratios differ between record and replay? Or that both record and replay use a mix of aspect ratios? And why do different aspect ratios result in this problem with the buddy allocator?}
%ORIGINALThe input to the \texttt{video\_processing} function contains a relatively large video fragment so that the pages for the video differ from an invocation to invocation (Fig.~\ref{fig:char_pages_reuse}). 
The orchestrator has to serve the missing pages one-by-one as page faults arise.
However, the end-to-end cold-start delay for \texttt{video\_processing} is nonetheless reduced as the longer function processing time is offset by faster re-connection to the function. 
We highlight, however, that functions with large inputs or control-flow that differs substantially across invocations may benefit less from REAP.

We repeat the same experiment in the presence of the invocation traffic to 20 warm, i.e., memory resident, functions and observe that the obtained data is within 5\% of Fig.~\ref{fig:eval_one_instance} results.
Also, we measure the efficacy of REAP on the same server but store the snapshots on a 2TB Western Digital WD2000F9YZ SATA3 7200 RPM HDD, instead of the SSD, and observed a 5.4$\times$ speed-up (not shown in the figure), on average, with REAP over baseline snapshotting.
%\marios{is this somewhere in the graphs?} \boris{If not, add "(not shown in the figure)" at the end of the sentence.}
}

\subsection{\tac{} Record Phase}
\label{sec:eval_record}
%\marios{are there any graphs for that? maybe add a small table? It's weird when data appears only on text}
\tac{} incurs a one-time overhead for recording the trace and the WS files.
%\boris{this says trace AND ws file. But Sec 4 makes it clear that only WS file needs to be recorded. We can't have these inconsistencies!}. 
Upon the first invocation of a function, this one-time overhead increases the end-to-end function invocation time by 15-87\% (28\% on average). Since most functions that we study have small dynamic inputs, they exhibit relatively small overheads of 12-34\%, with \texttt{image\_rotate} being an outlier with a performance degradation of 87\%. 

Because of the high speedups provided by \tac{} on all subsequent invocations of a function, and because the vast majority of functions execute multiple times~\cite{shahrad:serverless}, we conclude that \tac's one-time record overhead is easily amortized.
%Azure reports that >96\% of their functions are invoked at least twice per week (\S\ref{sec:serv_characteristics}).

%Fig.~\ref{fig:eval_one_instance} compares the cold-start delays of all the functions that we study with three configurations: the baseline Firecracker snapshots (\emph{snaps}), the REAP \emph{record} phase, which includes recording the function's trace and the contents of the corresponding working set pages, and REAP's \emph{prefetch} phase. One can see that albeit recording imposes a moderate XX-YY\% overhead (ZZ\% on average) for the first invocation, all further invocations become XX-YY$\times$ faster (2.7$\times$ on average). The time fraction that stands for reconnecting to the function's gRPC server shrinks by 20-30$\times$ to mere 6-7ms thanks to the stable subset of the function's working sets that accounts for this core serverless infrastructure functionality. The function processing latency component shows a significant reduction by 3.3$\times$ for all functions, except for \texttt{image\_rotate} that shows an increase in this latency component due to the significant number of pages that comprise a different image input to the function. The orchestrator has to serve these missing pages one-by-one as the page faults arise, leading to 2$\times$ slowdown. However, the end-to-end cold-start delay reduces even for \texttt{image\_rotate} where the function processing slowdown is offset by faster reconnecting to the function. 

\subsection{\tac{} Scalability}
\label{sec:eval_scalability}

\begin{figure}[t]
\centering
\includegraphics[width=0.47\textwidth]{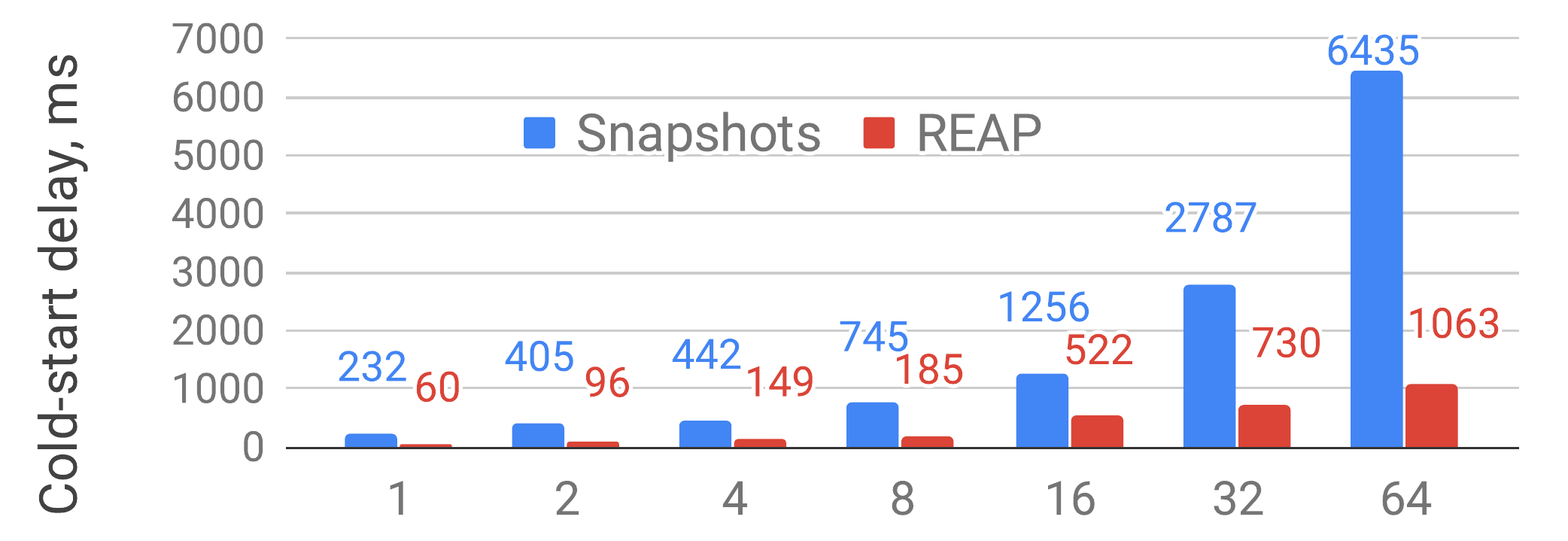}

\caption{Average instance cold-start delay while sweeping the number of the concurrently loading instances.
%\plamen{Add caption \boris{Is the Y-axis 'total delay' or 'per function'? Please clarify either in the axis label or in the caption.} }
}
\label{fig:eval_scalability}
\end{figure}

We demonstrate that \tac{} orchestrator retains its efficiency in the face of higher load. Specifically, we measure the average time that an instance takes to load from a snapshot and serve one function invocation when multiple independent functions arrive concurrently. We use the the \texttt{helloworld} function and consider up to 64 concurrent independent function arrivals. Fig.~\ref{fig:eval_scalability} shows the result of the study, comparing \tac{} to the baseline snapshots.

%\boris{PROBLEM: For 1 instance (1st bar), the baseline latency is shown as 259. This is inconsistent with previously-reported values of 232 (Fig 8 and others).} FIXED

Concurrently loading function instances should be able to take advantage of the multi-core CPU and abundant SSD bandwidth (48 logical cores and 850MB/s peak measured SSD bandwidth in our platform). Thus, we expect that as the degree of concurrency increases, the average per-instance latency will not significantly increase thanks to the available parallelism. Indeed, \tac{}'s cold invocation latency stays relatively low, increasing from 70ms to 185ms when the number of concurrent function arrivals goes from 1 to 8. 
By contrast, the baseline's per-instance invocation time shows a near-linear growth with the number of concurrently-arriving functions.
%, indicating that SSD throughput, which is limited with 4KB (page-sized) requests, is a bottleneck. 
We measure that the SSD throughput that the baseline instances are able to collectively extract is limited to mere 32MB/s for a single instance and 81MB/s for 64 concurrent instances.\footnote{We compute the SSD throughput per instance as the working set size divided by the average loading time.} Compared to the baseline, \tac{} is able to achieve 118-493MB/s, which explains its lower latency {\em and} better scalability.
%which allows sub-linear scaling of \tac{} invocation time up when booting up to 8 concurrent instances. 
Starting from the concurrency degree of 16, \tac{} becomes disk-bandwidth bound and its scalability is diminished.

\section{Discussion}

\cor{
\subsection{\tac's Efficiency and Mispredictions}
\tac's efficiency depends on how quickly the orchestrator can retrieve the guest memory pages from the snapshot storage and the percentage of the retrieved but unused pages. If the snapshots are located in a remote storage service (e.g., S3 or EBS), the retrieval speed depends on the amount of data to be moved and the latency and bandwidth of the network between the host and the service as well as the latency and bandwidth of the service’s internal disks. 

\tac{} reduces both the network and the disk bottlenecks by proactively moving a minimal amount of state.
However, \tac{} may fetch a modest number of pages that are not accessed during processing of some invocations. Our analysis shows that the fraction of mispredicted pages during a cold invocation is close to the “Unique” pages metric, shown in Fig.~\ref{fig:char_pages_reuse}, which is 3-39\%. These mispredictions have no impact on system correctness. The cost of these mispredictions is a modest increase in SSD bandwidth usage, proportionate to the fraction of the mispredicted pages.

\subsection{Applicability to Real-World Functions}
%Public cloud deployments impose strict economical and security requirements, meaning that the performance benefits of \tac{} should justify additional working set and trace files storage without undermining platform security.

Although \tac{} is applicable to the vast majority of functions, some functions may not benefit from \tac{}. For these functions, the additional working set and trace files may not be justified.
Prior work shows that 90\% of Azure functions are invoked less than once per minute, making these functions the primary target for \tac{}~\cite{shahrad:serverless}. Functions that are invoked very rarely (e.g., 3.5\% of functions are invoked less frequently than once per week) or more frequently than once per minute (and thus remaining warm) are unlikely to benefit from snapshot-based solutions.
Also, \tac{} is ill-suited for the functions where the first invocation is not representative of future invocations although we do not observe such behavior in our studies. In this pathological case, the orchestrator can easily detect low working set pages usage and either repeat \tac{}’s record phase or fall back to vanilla Firecracker snapshots for future invocations. For detection, the orchestrator could monitor the number of page faults that occur after the working set pages are installed, comparing this number to the number of pages in the working set.

\subsection{Security Concerns}
Similar to other snapshot techniques, spawning virtual machine clones from the same VM snapshot with \tac{} has implications for overall platform's security. In a naive snapshotting implementation, these VM clones may have an identical state for random number generators (i.e., poor entropy) and the same guest physical memory layout. The former problem may be addressed at the system level with hardware support for random number generation albeit the user-level random number generation libraries may remain vulnerable~\cite{firecracker:entropy,DBLP:conf/sp/EverspaughZJRS14}. The latter problem may lead to compromised ASLR, allowing the attacker to obtain the information about the guest memory layout. One mitigation strategy could be periodic re-generation of a snapshot (as well as the working set file and the trace file, for \tac{}). 
Alternatively, similarly to prior work on after-fork memory layout randomization~\cite{lu:how}, the orchestrator can dynamically re-randomize the guest memory placement while loading the VM's working set from the snapshot in the record phase of \tac{}. This would require modifying the guest page tables, with the hypervisor support, according to the new guest memory layout.

}
\section{Related Work}
\label{sec:related}

\cor{
\subsection{Open-Source Serverless Platforms}

Researchers release a number of benchmarks for serverless platforms. vHive adopts dockerized benchmarks from FunctionBench that provides a diverse set of Python benchmarks~\cite{kim:functionbench,kim:practical}. ServerlessBench contains a number of multi-function benchmarks, focusing on function composition patterns~\cite{yu:characterizing}. Researchers and practitioners release a range of systems that combine the FaaS programming model and autoscaling~\cite{apache:openwhisk,fission,fn_project,openlambda,kubeless}. Most of these platforms, however, rely on Docker or language sandboxes for isolating the untrusted user-provided function code that is often considered insufficiently secure in public cloud deployments~\cite{bianchini:socc_keynote,shahrad:serverless}. Kata Containers~\cite{kata:containers} and gVisor~\cite{google:gvisor} provide virtualized runtimes that are CRI-compliant but do not provide a toolchain for functions deployment and end-to-end evaluation and do not support snapshotting.
%As a result, the two leaders of serverless computing, namely Azure and AWS, choose virtual machines as the key function isolation technology. 
Compared to these systems, vHive is a open-source serverless experimentation platform -- representative of the production serverless platforms, like AWS Lambda~\cite{aws:lambdas} -- that uses latest virtualization, snapshotting, and cluster orchestration technologies combined with a framework for functions deployment and benchmarking.

\subsection{Virtual Machine Snapshots}
Originally, VM snapshots have been introduced for live migration before serverless computing emerged~\cite{clark:migration,nelson:migration}. The Potemkin project propose flash VM cloning to accelerate VM instantiation with copy-on-write memory sharing~\cite{vrable:scalability}. Snowflock extends the idea of VM cloning to instantiating VMs across entire clusters, relying on lazy guest memory loading to avoid large transfers of guest memory contents across network~\cite{lagar:snowflock}. To minimize the time spent in serving the series of lazy page faults during guest memory loading, the researchers explore a variety of working set prediction and prefetching techniques~\cite{zhang:fast,zhang:optimizing,zhu:twinkle,knauth:dreamserver}. These techniques rely on profiling of the memory accesses {\em after} the moment a checkpoint was taken and inspecting the locality characteristics of the guest OS' virtual address space.
Compared to these techniques, our work shows that serverless functions do not require complex working set estimation algorithms: it is sufficient to capture the pages that are accessed from the moment the vHive-CRI orchestrator forwards the invocation request to the function until the orchestrators receives the response from that function. Moreover, we find that extensive profiling may significantly bloat the captured working set, hence slowing down loading of future function instances, due to the guest OS activity that is not related to function processing.

}

\subsection{Serverless Cold-Start Optimizations}

% special hypervisors
Researchers have identified the problem of slow VM boot times, proposing solutions across the software stack to address it.
%\cor{The history saw a debate...~\cite{randal:ideal}}
Firecracker~\cite{agache:firecracker} and Cloud Hypervisor~\cite{cloud:hypervisor} use a specialized VMM that includes only the necessary functionality to run serverless workloads, while still running functions inside a full-blown, albeit minimal, Linux guest OS.
Dune~\cite{belay:dune} implements process-level virtualization.
Unikernels~\cite{cadden:seuss,kivity:osv,hand:unikernels,manco:my} leverage programming language techniques to aggressively perform deadcode elimination and create function-specific VM images, but sacrifice generality.
Finally, language sandboxes, e.g. Cloudflare Workers~\cite{cloudflare:workers} and Faasm~\cite{shillaker:faasm}, avoid the hardware virtualization costs and offer language level isolation through techniques such as V8 isolates~\cite{v8:isolates} and WebAssembly~\cite{wasm}.
Such approaches reduce the start-up costs, but limit the function implementation language choices while providing weaker isolation guarantees than VMs.
%and depend their isolation guarantees on the guarantees provided by the underlying language technology, thus increasing the attack surface.
\tac{} targets serverless workloads but remains agnostic to the hypervisor and the software that runs inside the VM.

% caching
Caching is another approach to reduce start-up latency.
Several proposals investigate the idea of keeping pre-warmed, pre-initialized execution environments in memory and ready to process requests.
Zygote~\cite{zygote} was introduced to accelerate the instantiation of Java applications by forking pre-initialized processes.
The zygote idea has been used for serverless platforms in SOCK~\cite{oakes:sock}, while SAND~\cite{akkus:sand} allows the reuse of pre-initialized sandboxes across function invocations.
These proposals, though, trade-off low memory utilization for better function invocation latencies.
\tac{} is able to deliver low invocation latencies without occupying extra memory resources when function instances are idle.
%manages to reduce memory consumption without sacrificing latency.

% origin of snapshots
Prior work uses VM snapshots for cold-start latency reduction, although snapshots have been initially introduced for live migration and VM cloning before serverless computing emerged~\cite{clark:migration, lagar:snowflock,nelson:migration}.
Both Firecracker~\cite{agache:firecracker, fc:snaps} and gVisor with its checkpoint-restore functionality~\cite{google:gvisor} support snapshoting.
The state-of-the-art snapshotting solution, called Catalyzer, improves on gVisor's VM offering three design options for fast VM restoration~\cite{du:catalyzer}.
Besides the "cold-boot" optimization discussed in \S\ref{sec:snaps101}, Catalyzer also proposes "warm-boot" and "sfork" optimizations that provide further performance improvements but require memory sharing across different VMs. In a production serverless deployment, memory sharing is considered insecure and is generally disallowed~\cite{agache:firecracker,aws:prod}.
%leverages this functionality and shows that VMM restoration can be achieved in 10s of milliseconds, but depends on memory sharing across different VMs.

%Replayable execution identifies that functions in language-based sandboxes use a small amount of memory pages when processing a function invocation, as compared to the function post-boot footprints~\cite{wang:replayable}, that is similar to our findings for virtualization-based sandboxes.
%As part of our analysis, we confirm a similar phenomenon in memory access patterns of virtualization-based sandboxes. 

Replayable execution aims to minimize the memory footprint and skip the lengthy code generation of the language-based sandboxes by taking a snapshot after thousands of function invocations, exploiting a similar observation as this work~-- that functions use a small number of memory pages when processing a function invocation~\cite{wang:replayable}.
However, when loading a new instance, their design relies on lazy paging similar to other snapshotting  techniques~\cite{fc:snaps,du:catalyzer}. 
%To accelerate cold invocations, Replayable execution proposes taking a snapshot after thousands of function invocations when the the language-based sandbox's JIT compiler generates all the necessary code, relying on lazy paging when restoring a function instance similar to Catalyzer~\cite{du:catalyzer}.
%which can then captured with a snapshot. 
%Similar to Catalyzer~\cite{du:catalyzer}, Replayable execution relies on lazy paging when restoring a function instance so that cold function invocations face the similar page-fault related overheads that we identify in \S\ref{sec:char_lat}.
In contrast, our work shows that the working set of the guest memory pages of virtualization-based sandboxes can be captured during the very first invocation, and that all future invocations can be accelerated by prefetching the stable memory working set into the guest memory.

\section{Conclusion}

Optimizing cold-start delays is key to improving serverless clients experience while maintaining serverless computing affordable. Our analysis identifies that the root cause for high cold-start delays is that the state-of-the-art solutions populate the guest memory on demand when restoring a function instance from a snapshot. 
This results in thousands of page faults, which must be served serially and significantly slow down a function invocation. We further find that functions exhibit a small working set of the guest memory pages that remains stable across different function invocations. Based on these insights, we present the \tac{} orchestrator that records a function's working set of pages, upon the first invocation of the function, and speeds up all further invocations of the same function by eagerly prefetching the working set of the function into the guest memory of a newly loaded function instance.

\section*{Acknowledgements}

%\cor{New section }
The authors thank the anonymous reviewers and the paper's shepherd, Professor James Bornholt,  as well as Professors Antonio Barbalace and James Larus, the members of the EASE lab at the University of Edinburgh, the members of the DCSL and the VLSC lab at EPFL for the fruitful discussions and for their valuable feedback on this work, Theodor Amariucai and Shyam Jesalpura for helping with the release of the vHive software. The authors would like to specially thank Firecracker and Containerd developers at Amazon Web Services, particularly Adrian Catangiu and Kazuyoshi Kato, for their guidance in setting up the Firecracker and Containerd components. This research was supported by the Arm Center of Excellence at the University of Edinburgh, Microsoft Swiss JRC TTL-MSR project at EPFL, and an IBM PhD fellowship.

\bibliographystyle{ACM-Reference-Format}
\bibliography{./bibcloud/gen-abbrev,dblp,ref}


\begin{thebibliography}{68}

%%% the Web address.

\ifx \showCODEN    \undefined \def \showCODEN     #1{\unskip}     \fi
\ifx \showDOI      \undefined \def \showDOI       #1{#1}\fi
\ifx \showISBNx    \undefined \def \showISBNx     #1{\unskip}     \fi
\ifx \showISBNxiii \undefined \def \showISBNxiii  #1{\unskip}     \fi
\ifx \showISSN     \undefined \def \showISSN      #1{\unskip}     \fi
\ifx \showLCCN     \undefined \def \showLCCN      #1{\unskip}     \fi
\ifx \shownote     \undefined \def \shownote      #1{#1}          \fi
\ifx \showarticletitle \undefined \def \showarticletitle #1{#1}   \fi
\ifx \showURL      \undefined \def \showURL       {\relax}        \fi
% invisible to TeX
\providecommand\bibfield[2]{#2}
\providecommand\bibinfo[2]{#2}
\providecommand\natexlab[1]{#1}
\providecommand\showeprint[2][]{arXiv:#2}

\bibitem[\protect\citeauthoryear{??}{clo}{[n.d.]}]%
        {cloud:hypervisor}
 \bibinfo{year}{[n.d.]}\natexlab{}.
\newblock \bibinfo{title}{{Cloud Hypervisor}}.
\newblock
\newblock
\newblock
\shownote{Available at \url{https://github.com/cloud-hypervisor}.}


\bibitem[\protect\citeauthoryear{??}{goo}{[n.d.]}]%
        {google:grpc}
 \bibinfo{year}{[n.d.]}\natexlab{}.
\newblock \bibinfo{title}{{gRPC: A High-Performance, Open Source Universal RPC
  Framework}}.
\newblock
\newblock
\newblock
\shownote{Available at \url{https://grpc.io}.}


\bibitem[\protect\citeauthoryear{??}{kat}{[n.d.]}]%
        {kata:containers}
 \bibinfo{year}{[n.d.]}\natexlab{}.
\newblock \bibinfo{title}{{Kata Containers}}.
\newblock
\newblock
\newblock
\shownote{Available at \url{https://katacontainers.io}.}


\bibitem[\protect\citeauthoryear{??}{was}{[n.d.]}]%
        {wasm}
 \bibinfo{year}{[n.d.]}\natexlab{}.
\newblock \bibinfo{title}{WebAssembly}.
\newblock
\newblock
\newblock
\shownote{Available at \url{https://webassembly.org}.}


\bibitem[\protect\citeauthoryear{Agache, Brooker, Iordache, Liguori,
  Neugebauer, Piwonka, and Popa}{Agache et~al\mbox{.}}{2020}]%
        {agache:firecracker}
\bibfield{author}{\bibinfo{person}{Alexandru Agache}, \bibinfo{person}{Marc
  Brooker}, \bibinfo{person}{Alexandra Iordache}, \bibinfo{person}{Anthony
  Liguori}, \bibinfo{person}{Rolf Neugebauer}, \bibinfo{person}{Phil Piwonka},
  {and} \bibinfo{person}{Diana-Maria Popa}.} \bibinfo{year}{2020}\natexlab{}.
\newblock \showarticletitle{{Firecracker: Lightweight Virtualization for
  Serverless Applications}}. In \bibinfo{booktitle}{\emph{Proceedings of the
  17th Symposium on Networked Systems Design and Implementation (NSDI)}}.
  \bibinfo{pages}{419--434}.
\newblock


\bibitem[\protect\citeauthoryear{Akkus, Chen, Rimac, Stein, Satzke, Beck,
  Aditya, and Hilt}{Akkus et~al\mbox{.}}{2018}]%
        {akkus:sand}
\bibfield{author}{\bibinfo{person}{Istemi~Ekin Akkus},
  \bibinfo{person}{Ruichuan Chen}, \bibinfo{person}{Ivica Rimac},
  \bibinfo{person}{Manuel Stein}, \bibinfo{person}{Klaus Satzke},
  \bibinfo{person}{Andre Beck}, \bibinfo{person}{Paarijaat Aditya}, {and}
  \bibinfo{person}{Volker Hilt}.} \bibinfo{year}{2018}\natexlab{}.
\newblock \showarticletitle{{SAND: Towards High-Performance Serverless
  Computing}}. In \bibinfo{booktitle}{\emph{Proceedings of the 2018 USENIX
  Annual Technical Conference (ATC)}}. \bibinfo{pages}{923--935}.
\newblock


\bibitem[\protect\citeauthoryear{{Amazon}}{{Amazon}}{[n.d.]}]%
        {firecracker:demo}
\bibfield{author}{\bibinfo{person}{{Amazon}}.}
  \bibinfo{year}{[n.d.]}\natexlab{}.
\newblock \bibinfo{title}{{A Demo Running 4000 Firecracker MicroVMs}}.
\newblock
\newblock
\newblock
\shownote{Available at
  \url{https://github.com/firecracker-microvm/firecracker-demo}.}


\bibitem[\protect\citeauthoryear{{Apache}}{{Apache}}{[n.d.]}]%
        {apache:openwhisk}
\bibfield{author}{\bibinfo{person}{{Apache}}.}
  \bibinfo{year}{[n.d.]}\natexlab{}.
\newblock \bibinfo{title}{{OpenWhisk}}.
\newblock
\newblock
\newblock
\shownote{Available at \url{https://openwhisk.apache.org/}.}


\bibitem[\protect\citeauthoryear{Authors}{Authors}{[n.d.]a}]%
        {fission}
\bibfield{author}{\bibinfo{person}{The~Fission Authors}.}
  \bibinfo{year}{[n.d.]}\natexlab{a}.
\newblock \bibinfo{title}{{Fission: Open Source, Kubernetes-Native Serverless
  Framework}}.
\newblock
\newblock
\newblock
\shownote{Available at \url{https://fission.io}.}


\bibitem[\protect\citeauthoryear{Authors}{Authors}{[n.d.]b}]%
        {fn_project}
\bibfield{author}{\bibinfo{person}{The Fn~Project Authors}.}
  \bibinfo{year}{[n.d.]}\natexlab{b}.
\newblock \bibinfo{title}{Fn Project}.
\newblock
\newblock
\newblock
\shownote{Available at \url{https://fnproject.io}.}


\bibitem[\protect\citeauthoryear{Authors}{Authors}{[n.d.]c}]%
        {istio}
\bibfield{author}{\bibinfo{person}{The~Istio Authors}.}
  \bibinfo{year}{[n.d.]}\natexlab{c}.
\newblock \bibinfo{title}{Istio}.
\newblock
\newblock
\newblock
\shownote{Available at \url{https://istio.io}.}


\bibitem[\protect\citeauthoryear{Authors}{Authors}{[n.d.]d}]%
        {knative}
\bibfield{author}{\bibinfo{person}{The~Knative Authors}.}
  \bibinfo{year}{[n.d.]}\natexlab{d}.
\newblock \bibinfo{title}{Knative}.
\newblock
\newblock
\newblock
\shownote{Available at \url{https://knative.dev}.}


\bibitem[\protect\citeauthoryear{{AWS re:Invent}}{{AWS re:Invent}}{2019}]%
        {aws:lambdas}
\bibfield{author}{\bibinfo{person}{{AWS re:Invent}}.}
  \bibinfo{year}{2019}\natexlab{}.
\newblock \bibinfo{title}{A Serverless Journey: {AWS Lambda} Under the Hood}.
\newblock
\newblock


\bibitem[\protect\citeauthoryear{Baidu}{Baidu}{[n.d.]}]%
        {baidu:kata}
\bibfield{author}{\bibinfo{person}{Baidu}.} \bibinfo{year}{[n.d.]}\natexlab{}.
\newblock \bibinfo{title}{The Application of {Kata Containers in Baidu AI
  Cloud}}.
\newblock
\newblock
\newblock
\shownote{Available at
  \url{https://katacontainers.io/collateral/ApplicationOfKataContainersInBaiduAICloud.pdf}.}


\bibitem[\protect\citeauthoryear{Belay, Bittau, Mashtizadeh, Terei, Mazières,
  and Kozyrakis}{Belay et~al\mbox{.}}{2012}]%
        {belay:dune}
\bibfield{author}{\bibinfo{person}{Adam Belay}, \bibinfo{person}{Andrea
  Bittau}, \bibinfo{person}{Ali~José Mashtizadeh}, \bibinfo{person}{David
  Terei}, \bibinfo{person}{David Mazières}, {and} \bibinfo{person}{Christos
  Kozyrakis}.} \bibinfo{year}{2012}\natexlab{}.
\newblock \showarticletitle{{Dune: Safe User-level Access to Privileged CPU
  Features}}. In \bibinfo{booktitle}{\emph{Proceedings of the 10th Symposium on
  Operating System Design and Implementation (OSDI)}}.
  \bibinfo{pages}{335--348}.
\newblock


\bibitem[\protect\citeauthoryear{Bianchini}{Bianchini}{[n.d.]}]%
        {bianchini:socc_keynote}
\bibfield{author}{\bibinfo{person}{Ricardo Bianchini}.}
  \bibinfo{year}{[n.d.]}\natexlab{}.
\newblock \bibinfo{title}{Serverless in Seattle: Toward Making Serverless the
  Future of the Cloud}.
\newblock
\newblock
\newblock
\shownote{Available at \url{https://acmsocc.github.io/2020/keynotes.html}.}


\bibitem[\protect\citeauthoryear{Cadden, Unger, Awad, Dong, Krieger, and
  Appavoo}{Cadden et~al\mbox{.}}{2020}]%
        {cadden:seuss}
\bibfield{author}{\bibinfo{person}{James Cadden}, \bibinfo{person}{Thomas
  Unger}, \bibinfo{person}{Yara Awad}, \bibinfo{person}{Han Dong},
  \bibinfo{person}{Orran Krieger}, {and} \bibinfo{person}{Jonathan Appavoo}.}
  \bibinfo{year}{2020}\natexlab{}.
\newblock \showarticletitle{{SEUSS: Skip Redundant Paths to Make Serverless
  Fast}}. In \bibinfo{booktitle}{\emph{Proceedings of the 2020 EuroSys
  Conference}}. \bibinfo{pages}{32:1--32:15}.
\newblock


\bibitem[\protect\citeauthoryear{{CBINSIGHTS}}{{CBINSIGHTS}}{[n.d.]}]%
        {market2}
\bibfield{author}{\bibinfo{person}{{CBINSIGHTS}}.}
  \bibinfo{year}{[n.d.]}\natexlab{}.
\newblock \bibinfo{title}{Why Serverless Computing Is The Fastest-Growing Cloud
  Services Segment}.
\newblock
\newblock
\newblock
\shownote{Available at
  \url{https://www.cbinsights.com/research/serverless-cloud-computing}.}


\bibitem[\protect\citeauthoryear{Clark, Fraser, Hand, Hansen, Jul, Limpach,
  Pratt, and Warfield}{Clark et~al\mbox{.}}{2005}]%
        {clark:migration}
\bibfield{author}{\bibinfo{person}{Christopher Clark}, \bibinfo{person}{Keir
  Fraser}, \bibinfo{person}{Steven Hand}, \bibinfo{person}{Jacob~Gorm Hansen},
  \bibinfo{person}{Eric Jul}, \bibinfo{person}{Christian Limpach},
  \bibinfo{person}{Ian Pratt}, {and} \bibinfo{person}{Andrew Warfield}.}
  \bibinfo{year}{2005}\natexlab{}.
\newblock \showarticletitle{{Live Migration of Virtual Machines}}. In
  \bibinfo{booktitle}{\emph{Proceedings of the 2nd Symposium on Networked
  Systems Design and Implementation (NSDI)}}.
\newblock


\bibitem[\protect\citeauthoryear{{Cloud Native Computing Foundation}}{{Cloud
  Native Computing Foundation}}{[n.d.]}]%
        {cri-o}
\bibfield{author}{\bibinfo{person}{{Cloud Native Computing Foundation}}.}
  \bibinfo{year}{[n.d.]}\natexlab{}.
\newblock \bibinfo{title}{{CRI-O}: Lightweight Container Runtime for
  Kubernetes}.
\newblock
\newblock
\newblock
\shownote{Available at \url{https://cri-o.io}.}


\bibitem[\protect\citeauthoryear{{Containerd}}{{Containerd}}{[n.d.]}]%
        {containerd:industry}
\bibfield{author}{\bibinfo{person}{{Containerd}}.}
  \bibinfo{year}{[n.d.]}\natexlab{}.
\newblock \bibinfo{title}{{An Industry-Standard Container Runtime with an
  Emphasis on Simplicity, Robustness and Portability}}.
\newblock
\newblock
\newblock
\shownote{Available at \url{https://containerd.io}.}


\bibitem[\protect\citeauthoryear{{CouldFlare}}{{CouldFlare}}{[n.d.]}]%
        {cloudflare:workers}
\bibfield{author}{\bibinfo{person}{{CouldFlare}}.}
  \bibinfo{year}{[n.d.]}\natexlab{}.
\newblock \bibinfo{title}{CloudFlare Workers}.
\newblock
\newblock
\newblock
\shownote{Available at \url{https://workers.cloudflare.com/}.}


\bibitem[\protect\citeauthoryear{{Daniel Krook}}{{Daniel Krook}}{[n.d.]}]%
        {krook:openwhisk}
\bibfield{author}{\bibinfo{person}{{Daniel Krook}}.}
  \bibinfo{year}{[n.d.]}\natexlab{}.
\newblock \bibinfo{title}{Five Minute Intro to Open Source Serverless
  Development with {OpenWhisk}}.
\newblock
\newblock
\newblock
\shownote{Available at
  \url{https://medium.com/openwhisk/five-minute-intro-to-open-source-serverless-development-with-openwhisk-328b0ebfa160}.}


\bibitem[\protect\citeauthoryear{Developers}{Developers}{[n.d.]}]%
        {zygote}
\bibfield{author}{\bibinfo{person}{Android Developers}.}
  \bibinfo{year}{[n.d.]}\natexlab{}.
\newblock \bibinfo{title}{Overview of Memory Management}.
\newblock
\newblock
\newblock
\shownote{Available at
  \url{https://developer.android.com/topic/performance/memory-overview}.}


\bibitem[\protect\citeauthoryear{Docker}{Docker}{[n.d.]}]%
        {docker:devmapper}
\bibfield{author}{\bibinfo{person}{Docker}.} \bibinfo{year}{[n.d.]}\natexlab{}.
\newblock \bibinfo{title}{Use the Device Mapper Storage Driver}.
\newblock
\newblock
\newblock
\shownote{Available at
  \url{https://docs.docker.com/storage/storagedriver/device-mapper-driver}.}


\bibitem[\protect\citeauthoryear{Du, Yu, Xia, Zang, Yan, Qin, Wu, and Chen}{Du
  et~al\mbox{.}}{2020}]%
        {du:catalyzer}
\bibfield{author}{\bibinfo{person}{Dong Du}, \bibinfo{person}{Tianyi Yu},
  \bibinfo{person}{Yubin Xia}, \bibinfo{person}{Binyu Zang},
  \bibinfo{person}{Guanglu Yan}, \bibinfo{person}{Chenggang Qin},
  \bibinfo{person}{Qixuan Wu}, {and} \bibinfo{person}{Haibo Chen}.}
  \bibinfo{year}{2020}\natexlab{}.
\newblock \showarticletitle{{Catalyzer: Sub-Millisecond Startup for Serverless
  Computing with Initialization-less Booting}}. In
  \bibinfo{booktitle}{\emph{Proceedings of the 25th International Conference on
  Architectural Support for Programming Languages and Operating Systems
  (ASPLOS-XXV)}}. \bibinfo{pages}{467--481}.
\newblock


\bibitem[\protect\citeauthoryear{Everspaugh, Zhai, Jellinek, Ristenpart, and
  Swift}{Everspaugh et~al\mbox{.}}{2014}]%
        {DBLP:conf/sp/EverspaughZJRS14}
\bibfield{author}{\bibinfo{person}{Adam Everspaugh}, \bibinfo{person}{Yan
  Zhai}, \bibinfo{person}{Robert Jellinek}, \bibinfo{person}{Thomas
  Ristenpart}, {and} \bibinfo{person}{Michael~M. Swift}.}
  \bibinfo{year}{2014}\natexlab{}.
\newblock \showarticletitle{Not-So-Random Numbers in Virtualized Linux and the
  Whirlwind {RNG}}. In \bibinfo{booktitle}{\emph{Proceedings of the 35th IEEE
  Symposium on Security and Privacy (S\&P)}}. \bibinfo{pages}{559--574}.
\newblock


\bibitem[\protect\citeauthoryear{Google}{Google}{[n.d.]}]%
        {google:gvisor}
\bibfield{author}{\bibinfo{person}{Google}.} \bibinfo{year}{[n.d.]}\natexlab{}.
\newblock \bibinfo{title}{{gVisor}}.
\newblock
\newblock
\newblock
\shownote{Available at \url{https://gvisor.dev}.}


\bibitem[\protect\citeauthoryear{{Google Cloud}}{{Google Cloud}}{[n.d.]}]%
        {google:warm_req}
\bibfield{author}{\bibinfo{person}{{Google Cloud}}.}
  \bibinfo{year}{[n.d.]}\natexlab{}.
\newblock \bibinfo{title}{Configuring Warmup Requests to Improve Performance}.
\newblock
\newblock
\newblock
\shownote{Available at
  \url{https://cloud.google.com/appengine/docs/standard/python/configuring-warmup-requests}.}


\bibitem[\protect\citeauthoryear{Hendrickson, Sturdevant, Harter,
  Venkataramani, Arpaci{-}Dusseau, and Arpaci{-}Dusseau}{Hendrickson
  et~al\mbox{.}}{2016}]%
        {openlambda}
\bibfield{author}{\bibinfo{person}{Scott Hendrickson}, \bibinfo{person}{Stephen
  Sturdevant}, \bibinfo{person}{Tyler Harter}, \bibinfo{person}{Venkateshwaran
  Venkataramani}, \bibinfo{person}{Andrea~C. Arpaci{-}Dusseau}, {and}
  \bibinfo{person}{Remzi~H. Arpaci{-}Dusseau}.}
  \bibinfo{year}{2016}\natexlab{}.
\newblock \showarticletitle{Serverless Computation with {OpenLambda}}. In
  \bibinfo{booktitle}{\emph{8th {USENIX} Workshop on Hot Topics in Cloud
  Computing (HotCloud)}}.
\newblock


\bibitem[\protect\citeauthoryear{Kaffes, Yadwadkar, and Kozyrakis}{Kaffes
  et~al\mbox{.}}{2019}]%
        {kaffes:centralized}
\bibfield{author}{\bibinfo{person}{Kostis Kaffes}, \bibinfo{person}{Neeraja~J.
  Yadwadkar}, {and} \bibinfo{person}{Christos Kozyrakis}.}
  \bibinfo{year}{2019}\natexlab{}.
\newblock \showarticletitle{Centralized Core-Granular Scheduling for Serverless
  Functions}. In \bibinfo{booktitle}{\emph{Proceedings of the 2019 ACM
  Symposium on Cloud Computing (SOCC)}}. \bibinfo{pages}{158--164}.
\newblock


\bibitem[\protect\citeauthoryear{Kim and Lee}{Kim and Lee}{2019a}]%
        {kim:functionbench}
\bibfield{author}{\bibinfo{person}{Jeongchul Kim} {and}
  \bibinfo{person}{Kyungyong Lee}.} \bibinfo{year}{2019}\natexlab{a}.
\newblock \showarticletitle{{FunctionBench: A Suite of Workloads for Serverless
  Cloud Function Service}}. In \bibinfo{booktitle}{\emph{Proceedings of the
  12th IEEE International Conference on Cloud Computing (CLOUD)}}.
  \bibinfo{pages}{502--504}.
\newblock


\bibitem[\protect\citeauthoryear{Kim and Lee}{Kim and Lee}{2019b}]%
        {kim:practical}
\bibfield{author}{\bibinfo{person}{Jeongchul Kim} {and}
  \bibinfo{person}{Kyungyong Lee}.} \bibinfo{year}{2019}\natexlab{b}.
\newblock \showarticletitle{{Practical Cloud Workloads for Serverless FaaS}}.
  In \bibinfo{booktitle}{\emph{Proceedings of the 2019 ACM Symposium on Cloud
  Computing (SOCC)}}. \bibinfo{pages}{477}.
\newblock


\bibitem[\protect\citeauthoryear{Kivity, Laor, Costa, Enberg, Har'El, Marti,
  and Zolotarov}{Kivity et~al\mbox{.}}{2014}]%
        {kivity:osv}
\bibfield{author}{\bibinfo{person}{Avi Kivity}, \bibinfo{person}{Dor Laor},
  \bibinfo{person}{Glauber Costa}, \bibinfo{person}{Pekka Enberg},
  \bibinfo{person}{Nadav Har'El}, \bibinfo{person}{Don Marti}, {and}
  \bibinfo{person}{Vlad Zolotarov}.} \bibinfo{year}{2014}\natexlab{}.
\newblock \showarticletitle{{OSv - Optimizing the Operating System for Virtual
  Machines}}. In \bibinfo{booktitle}{\emph{Proceedings of the 2014 USENIX
  Annual Technical Conference (ATC)}}. \bibinfo{pages}{61--72}.
\newblock


\bibitem[\protect\citeauthoryear{Knauth and Fetzer}{Knauth and Fetzer}{2014}]%
        {knauth:dreamserver}
\bibfield{author}{\bibinfo{person}{Thomas Knauth} {and}
  \bibinfo{person}{Christof Fetzer}.} \bibinfo{year}{2014}\natexlab{}.
\newblock \showarticletitle{{DreamServer: Truly On-Demand Cloud Services}}. In
  \bibinfo{booktitle}{\emph{Proceedings of the 7th ACM International Systems
  and Storage Conference (SYSTOR)}}. \bibinfo{pages}{9:1--9:11}.
\newblock


\bibitem[\protect\citeauthoryear{Kubeless}{Kubeless}{[n.d.]}]%
        {kubeless}
\bibfield{author}{\bibinfo{person}{Kubeless}.}
  \bibinfo{year}{[n.d.]}\natexlab{}.
\newblock \bibinfo{title}{Kubeless: The Kubernetes Native Serverless
  Framework}.
\newblock
\newblock
\newblock
\shownote{Available at \url{https://kubeless.io}.}


\bibitem[\protect\citeauthoryear{{Kubernetes}}{{Kubernetes}}{[n.d.]}]%
        {k8s}
\bibfield{author}{\bibinfo{person}{{Kubernetes}}.}
  \bibinfo{year}{[n.d.]}\natexlab{}.
\newblock \bibinfo{title}{Production-Grade Container Orchestration}.
\newblock
\newblock
\newblock
\shownote{Available at \url{https://kubernetes.io}.}


\bibitem[\protect\citeauthoryear{Lagar-Cavilla, Whitney, Scannell, Patchin,
  Rumble, de~Lara, Brudno, and Satyanarayanan}{Lagar-Cavilla
  et~al\mbox{.}}{2009}]%
        {lagar:snowflock}
\bibfield{author}{\bibinfo{person}{Horacio~Andrés Lagar-Cavilla},
  \bibinfo{person}{Joseph~Andrew Whitney}, \bibinfo{person}{Adin~Matthew
  Scannell}, \bibinfo{person}{Philip Patchin}, \bibinfo{person}{Stephen~M.
  Rumble}, \bibinfo{person}{Eyal de Lara}, \bibinfo{person}{Michael Brudno},
  {and} \bibinfo{person}{Mahadev Satyanarayanan}.}
  \bibinfo{year}{2009}\natexlab{}.
\newblock \showarticletitle{{SnowFlock: rapid virtual machine cloning for cloud
  computing}}. In \bibinfo{booktitle}{\emph{Proceedings of the 2009 EuroSys
  Conference}}. \bibinfo{pages}{1--12}.
\newblock


\bibitem[\protect\citeauthoryear{{Linux programmer's manual}}{{Linux
  programmer's manual}}{[n.d.]}]%
        {man:userfaultfd}
\bibfield{author}{\bibinfo{person}{{Linux programmer's manual}}.}
  \bibinfo{year}{[n.d.]}\natexlab{}.
\newblock \bibinfo{title}{Userfaultfd}.
\newblock
\newblock
\newblock
\shownote{Available at
  \url{https://man7.org/linux/man-pages/man2/userfaultfd.2.html}.}


\bibitem[\protect\citeauthoryear{Lu, Lee, Nürnberger, and Backes}{Lu
  et~al\mbox{.}}{2016}]%
        {lu:how}
\bibfield{author}{\bibinfo{person}{Kangjie Lu}, \bibinfo{person}{Wenke Lee},
  \bibinfo{person}{Stefan Nürnberger}, {and} \bibinfo{person}{Michael
  Backes}.} \bibinfo{year}{2016}\natexlab{}.
\newblock \showarticletitle{{How to Make ASLR Win the Clone Wars: Runtime
  Re-Randomization}}. In \bibinfo{booktitle}{\emph{Proceedings of the 2016
  Annual Network and Distributed System Security Symposium (NDSS)}}.
\newblock


\bibitem[\protect\citeauthoryear{Madhavapeddy, Mortier, Rotsos, Scott, Singh,
  Gazagnaire, Smith, Hand, and Crowcroft}{Madhavapeddy et~al\mbox{.}}{2013}]%
        {hand:unikernels}
\bibfield{author}{\bibinfo{person}{Anil Madhavapeddy}, \bibinfo{person}{Richard
  Mortier}, \bibinfo{person}{Charalampos Rotsos}, \bibinfo{person}{David~J.
  Scott}, \bibinfo{person}{Balraj Singh}, \bibinfo{person}{Thomas Gazagnaire},
  \bibinfo{person}{Steven Smith}, \bibinfo{person}{Steven Hand}, {and}
  \bibinfo{person}{Jon Crowcroft}.} \bibinfo{year}{2013}\natexlab{}.
\newblock \showarticletitle{Unikernels: Library Operating Systems for the
  Cloud}. In \bibinfo{booktitle}{\emph{Proceedings of the 18th International
  Conference on Architectural Support for Programming Languages and Operating
  Systems (ASPLOS-XVIII)}}. \bibinfo{pages}{461--472}.
\newblock


\bibitem[\protect\citeauthoryear{man page}{man page}{[n.d.]}]%
        {fio}
\bibfield{author}{\bibinfo{person}{Linux man page}.}
  \bibinfo{year}{[n.d.]}\natexlab{}.
\newblock \bibinfo{title}{fio}.
\newblock
\newblock
\newblock
\shownote{Available at \url{https://linux.die.net/man/1/fio}.}


\bibitem[\protect\citeauthoryear{Manco, Lupu, Schmidt, Mendes, Kuenzer, Sati,
  Yasukata, Raiciu, and Huici}{Manco et~al\mbox{.}}{2017}]%
        {manco:my}
\bibfield{author}{\bibinfo{person}{Filipe Manco}, \bibinfo{person}{Costin
  Lupu}, \bibinfo{person}{Florian Schmidt}, \bibinfo{person}{Jose Mendes},
  \bibinfo{person}{Simon Kuenzer}, \bibinfo{person}{Sumit Sati},
  \bibinfo{person}{Kenichi Yasukata}, \bibinfo{person}{Costin Raiciu}, {and}
  \bibinfo{person}{Felipe Huici}.} \bibinfo{year}{2017}\natexlab{}.
\newblock \showarticletitle{{My VM is Lighter (and Safer) than your
  Container}}. In \bibinfo{booktitle}{\emph{Proceedings of the 26th ACM
  Symposium on Operating Systems Principles (SOSP)}}.
  \bibinfo{pages}{218--233}.
\newblock


\bibitem[\protect\citeauthoryear{{Market Reports World}}{{Market Reports
  World}}{2019}]%
        {market1}
\bibfield{author}{\bibinfo{person}{{Market Reports World}}.}
  \bibinfo{year}{2019}\natexlab{}.
\newblock \bibinfo{title}{{Serverless Architecture Market by End-Users and
  Geography - Global Forecast 2019-2023}}.
\newblock
\newblock
\newblock
\shownote{Available at
  \url{https://www.marketreportsworld.com/serverless-architecture-market-13684687}.}


\bibitem[\protect\citeauthoryear{Microsoft}{Microsoft}{2019}]%
        {azure:functions}
\bibfield{author}{\bibinfo{person}{Microsoft}.}
  \bibinfo{year}{2019}\natexlab{}.
\newblock \bibinfo{title}{Azure Functions}.
\newblock
\newblock
\newblock
\shownote{Available at
  \url{https://azure.microsoft.com/en-gb/services/functions}.}


\bibitem[\protect\citeauthoryear{{MinIO}}{{MinIO}}{[n.d.]}]%
        {minio}
\bibfield{author}{\bibinfo{person}{{MinIO}}.}
  \bibinfo{year}{[n.d.]}\natexlab{}.
\newblock \bibinfo{title}{Kubernetes Native, High Performance Object Storage}.
\newblock
\newblock
\newblock
\shownote{Available at \url{https://min.io}.}


\bibitem[\protect\citeauthoryear{Nelson, Lim, and Hutchins}{Nelson
  et~al\mbox{.}}{2005}]%
        {nelson:migration}
\bibfield{author}{\bibinfo{person}{Michael Nelson}, \bibinfo{person}{Beng-Hong
  Lim}, {and} \bibinfo{person}{Greg Hutchins}.}
  \bibinfo{year}{2005}\natexlab{}.
\newblock \showarticletitle{{Fast Transparent Migration for Virtual Machines}}.
  In \bibinfo{booktitle}{\emph{USENIX Annual Technical Conference}}.
  \bibinfo{pages}{391--394}.
\newblock


\bibitem[\protect\citeauthoryear{Neves}{Neves}{[n.d.]}]%
        {warm_func}
\bibfield{author}{\bibinfo{person}{Goncalo Neves}.}
  \bibinfo{year}{[n.d.]}\natexlab{}.
\newblock \bibinfo{title}{Keeping Functions Warm -- How To Fix {AWS Lambda}
  Cold Start Issues}.
\newblock
\newblock
\newblock
\shownote{Available at
  \url{https://serverless.com/blog/keep-your-lambdas-warm}.}


\bibitem[\protect\citeauthoryear{Oakes, Yang, Zhou, Houck, Harter,
  Arpaci-Dusseau, and Arpaci-Dusseau}{Oakes et~al\mbox{.}}{2018}]%
        {oakes:sock}
\bibfield{author}{\bibinfo{person}{Edward Oakes}, \bibinfo{person}{Leon Yang},
  \bibinfo{person}{Dennis Zhou}, \bibinfo{person}{Kevin Houck},
  \bibinfo{person}{Tyler Harter}, \bibinfo{person}{Andrea~C. Arpaci-Dusseau},
  {and} \bibinfo{person}{Remzi~H. Arpaci-Dusseau}.}
  \bibinfo{year}{2018}\natexlab{}.
\newblock \showarticletitle{{SOCK: Rapid Task Provisioning with
  Serverless-Optimized Containers}}. In \bibinfo{booktitle}{\emph{Proceedings
  of the 2018 USENIX Annual Technical Conference (ATC)}}.
  \bibinfo{pages}{57--70}.
\newblock


\bibitem[\protect\citeauthoryear{{OpenNebula}}{{OpenNebula}}{[n.d.]}]%
        {opennebula:firework}
\bibfield{author}{\bibinfo{person}{{OpenNebula}}.}
  \bibinfo{year}{[n.d.]}\natexlab{}.
\newblock \bibinfo{title}{{OpenNebula + Firecracker: Building} the Future of
  On-Premises Serverless Computing}.
\newblock
\newblock
\newblock
\shownote{Available at
  \url{https://opennebula.io/opennebula-firecracker-building-the-future-of-on-premises-serverless-computing}.}


\bibitem[\protect\citeauthoryear{Randal}{Randal}{2020}]%
        {randal:ideal}
\bibfield{author}{\bibinfo{person}{Allison Randal}.}
  \bibinfo{year}{2020}\natexlab{}.
\newblock \showarticletitle{{The Ideal Versus the Real: Revisiting the History
  of Virtual Machines and Containers}}.
\newblock \bibinfo{journal}{\emph{ACM Comput. Surv.}} \bibinfo{volume}{53},
  \bibinfo{number}{1} (\bibinfo{year}{2020}), \bibinfo{pages}{5:1--5:31}.
\newblock


\bibitem[\protect\citeauthoryear{{Samuel Karp}}{{Samuel Karp}}{[n.d.]}]%
        {fc-ctrd:deep-dive}
\bibfield{author}{\bibinfo{person}{{Samuel Karp}}.}
  \bibinfo{year}{[n.d.]}\natexlab{}.
\newblock \bibinfo{title}{Deep Dive into Firecracker-Containerd}.
\newblock
\newblock
\newblock
\shownote{Available at
  \url{https://speakerdeck.com/samuelkarp/deep-dive-into-firecracker-containerd-re-invent-2019-con408}.}


\bibitem[\protect\citeauthoryear{Shahrad, Fonseca, Goiri, Chaudhry, Batum,
  Cooke, Laureano, Tresness, Russinovich, and Bianchini}{Shahrad
  et~al\mbox{.}}{2020}]%
        {shahrad:serverless}
\bibfield{author}{\bibinfo{person}{Mohammad Shahrad}, \bibinfo{person}{Rodrigo
  Fonseca}, \bibinfo{person}{Iñigo Goiri}, \bibinfo{person}{Gohar Chaudhry},
  \bibinfo{person}{Paul Batum}, \bibinfo{person}{Jason Cooke},
  \bibinfo{person}{Eduardo Laureano}, \bibinfo{person}{Colby Tresness},
  \bibinfo{person}{Mark Russinovich}, {and} \bibinfo{person}{Ricardo
  Bianchini}.} \bibinfo{year}{2020}\natexlab{}.
\newblock \showarticletitle{{Serverless in the Wild: Characterizing and
  Optimizing the Serverless Workload at a Large Cloud Provider}}. In
  \bibinfo{booktitle}{\emph{Proceedings of the 2020 USENIX Annual Technical
  Conference (ATC)}}. \bibinfo{pages}{205--218}.
\newblock


\bibitem[\protect\citeauthoryear{Shilkov}{Shilkov}{[n.d.]}]%
        {cold-start-war}
\bibfield{author}{\bibinfo{person}{Mikhail Shilkov}.}
  \bibinfo{year}{[n.d.]}\natexlab{}.
\newblock \bibinfo{title}{Serverless: Cold Start War}.
\newblock
\newblock
\newblock
\shownote{Available at
  \url{https://mikhail.io/2018/08/serverless-cold-start-war}.}


\bibitem[\protect\citeauthoryear{Shillaker and Pietzuch}{Shillaker and
  Pietzuch}{2020}]%
        {shillaker:faasm}
\bibfield{author}{\bibinfo{person}{Simon Shillaker} {and}
  \bibinfo{person}{Peter~R. Pietzuch}.} \bibinfo{year}{2020}\natexlab{}.
\newblock \showarticletitle{{Faasm: Lightweight Isolation for Efficient
  Stateful Serverless Computing}}. In \bibinfo{booktitle}{\emph{Proceedings of
  the 2020 USENIX Annual Technical Conference (ATC)}}.
  \bibinfo{pages}{419--433}.
\newblock


\bibitem[\protect\citeauthoryear{Strehl}{Strehl}{[n.d.]}]%
        {serverless:benchmark}
\bibfield{author}{\bibinfo{person}{Bernd Strehl}.}
  \bibinfo{year}{[n.d.]}\natexlab{}.
\newblock \bibinfo{title}{Lambda Serverless Benchmark}.
\newblock
\newblock
\newblock
\shownote{Available at \url{https://serverless-benchmark.com}.}


\bibitem[\protect\citeauthoryear{{The Firecracker Authors}}{{The Firecracker
  Authors}}{[n.d.]a}]%
        {firecracker:entropy}
\bibfield{author}{\bibinfo{person}{{The Firecracker Authors}}.}
  \bibinfo{year}{[n.d.]}\natexlab{a}.
\newblock \bibinfo{title}{{Entropy for Clones}}.
\newblock
\newblock
\newblock
\shownote{Available at
  \url{https://github.com/firecracker-microvm/firecracker/blob/master/docs/snapshotting/random-for-clones.md}.}


\bibitem[\protect\citeauthoryear{{The Firecracker Authors}}{{The Firecracker
  Authors}}{[n.d.]b}]%
        {fc:snaps}
\bibfield{author}{\bibinfo{person}{{The Firecracker Authors}}.}
  \bibinfo{year}{[n.d.]}\natexlab{b}.
\newblock \bibinfo{title}{Firecracker Snapshotting}.
\newblock
\newblock
\newblock
\shownote{Available at
  \url{https://github.com/firecracker-microvm/firecracker/blob/master/docs/snapshotting/snapshot-support.md}.}


\bibitem[\protect\citeauthoryear{{The Firecracker Authors}}{{The Firecracker
  Authors}}{[n.d.]c}]%
        {aws:prod}
\bibfield{author}{\bibinfo{person}{{The Firecracker Authors}}.}
  \bibinfo{year}{[n.d.]}\natexlab{c}.
\newblock \bibinfo{title}{Production Host Setup Recommendations}.
\newblock
\newblock
\newblock
\shownote{Available at
  \url{https://github.com/firecracker-microvm/firecracker/blob/master/docs/prod-host-setup.md}.}


\bibitem[\protect\citeauthoryear{{The Firecracker-Containerd Authors}}{{The
  Firecracker-Containerd Authors}}{[n.d.]}]%
        {fc-ctrd:github}
\bibfield{author}{\bibinfo{person}{{The Firecracker-Containerd Authors}}.}
  \bibinfo{year}{[n.d.]}\natexlab{}.
\newblock \bibinfo{title}{Firecracker-Containerd}.
\newblock
\newblock
\newblock
\shownote{Available at
  \url{https://github.com/firecracker-microvm/firecracker-containerd}.}


\bibitem[\protect\citeauthoryear{{The Linux Foundation Projects}}{{The Linux
  Foundation Projects}}{[n.d.]}]%
        {opencontainers}
\bibfield{author}{\bibinfo{person}{{The Linux Foundation Projects}}.}
  \bibinfo{year}{[n.d.]}\natexlab{}.
\newblock \bibinfo{title}{{Open Container Initiative}}.
\newblock
\newblock
\newblock
\shownote{Available at \url{https://opencontainers.org}.}


\bibitem[\protect\citeauthoryear{{V8}}{{V8}}{[n.d.]}]%
        {v8:isolates}
\bibfield{author}{\bibinfo{person}{{V8}}.} \bibinfo{year}{[n.d.]}\natexlab{}.
\newblock \bibinfo{title}{Isolate Class Reference}.
\newblock
\newblock
\newblock
\shownote{Available at
  \url{https://v8docs.nodesource.com/node-0.8/d5/dda/classv8_1_1_isolate.html}.}


\bibitem[\protect\citeauthoryear{Vrable, Ma, Chen, Moore, Vandekieft, Snoeren,
  Voelker, and Savage}{Vrable et~al\mbox{.}}{2005}]%
        {vrable:scalability}
\bibfield{author}{\bibinfo{person}{Michael Vrable}, \bibinfo{person}{Justin
  Ma}, \bibinfo{person}{Jay Chen}, \bibinfo{person}{David Moore},
  \bibinfo{person}{Erik Vandekieft}, \bibinfo{person}{Alex~C. Snoeren},
  \bibinfo{person}{Geoffrey~M. Voelker}, {and} \bibinfo{person}{Stefan
  Savage}.} \bibinfo{year}{2005}\natexlab{}.
\newblock \showarticletitle{{Scalability, Fidelity, and Containment in the
  Potemkin Virtual Honeyfarm}}. In \bibinfo{booktitle}{\emph{Proceedings of the
  20th ACM Symposium on Operating Systems Principles (SOSP)}}.
  \bibinfo{pages}{148--162}.
\newblock


\bibitem[\protect\citeauthoryear{Wang, Ho, and Wu}{Wang et~al\mbox{.}}{2019}]%
        {wang:replayable}
\bibfield{author}{\bibinfo{person}{Kai-Ting~Amy Wang}, \bibinfo{person}{Rayson
  Ho}, {and} \bibinfo{person}{Peng Wu}.} \bibinfo{year}{2019}\natexlab{}.
\newblock \showarticletitle{{Replayable Execution Optimized for Page Sharing
  for a Managed Runtime Environment}}. In \bibinfo{booktitle}{\emph{Proceedings
  of the 2019 EuroSys Conference}}. \bibinfo{pages}{39:1--39:16}.
\newblock


\bibitem[\protect\citeauthoryear{Yu, Liu, Du, Xia, Zang, Lu, Yang, Qin, and
  Chen}{Yu et~al\mbox{.}}{2020}]%
        {yu:characterizing}
\bibfield{author}{\bibinfo{person}{Tianyi Yu}, \bibinfo{person}{Qingyuan Liu},
  \bibinfo{person}{Dong Du}, \bibinfo{person}{Yubin Xia},
  \bibinfo{person}{Binyu Zang}, \bibinfo{person}{Ziqian Lu},
  \bibinfo{person}{Pingchao Yang}, \bibinfo{person}{Chenggang Qin}, {and}
  \bibinfo{person}{Haibo Chen}.} \bibinfo{year}{2020}\natexlab{}.
\newblock \showarticletitle{{Characterizing Serverless Platforms with
  ServerlessBench}}. In \bibinfo{booktitle}{\emph{Proceedings of the 2020 ACM
  Symposium on Cloud Computing (SOCC)}}. \bibinfo{pages}{30--44}.
\newblock


\bibitem[\protect\citeauthoryear{Zhang, Denniston, Baskakov, and
  Garthwaite}{Zhang et~al\mbox{.}}{2013}]%
        {zhang:optimizing}
\bibfield{author}{\bibinfo{person}{Irene Zhang}, \bibinfo{person}{Tyler
  Denniston}, \bibinfo{person}{Yury Baskakov}, {and} \bibinfo{person}{Alex
  Garthwaite}.} \bibinfo{year}{2013}\natexlab{}.
\newblock \showarticletitle{{Optimizing VM Checkpointing for Restore
  Performance in VMware ESXi}}. In \bibinfo{booktitle}{\emph{Proceedings of the
  2013 USENIX Annual Technical Conference (ATC)}}. \bibinfo{pages}{1--12}.
\newblock


\bibitem[\protect\citeauthoryear{Zhang, Garthwaite, Baskakov, and Barr}{Zhang
  et~al\mbox{.}}{2011}]%
        {zhang:fast}
\bibfield{author}{\bibinfo{person}{Irene Zhang}, \bibinfo{person}{Alex
  Garthwaite}, \bibinfo{person}{Yury Baskakov}, {and}
  \bibinfo{person}{Kenneth~C. Barr}.} \bibinfo{year}{2011}\natexlab{}.
\newblock \showarticletitle{{Fast Restore of Checkpointed Memory using Working
  Set Estimation}}. In \bibinfo{booktitle}{\emph{Proceedings of the 7th
  International Conference on Virtual Execution Environments (VEE)}}.
  \bibinfo{pages}{87--98}.
\newblock


\bibitem[\protect\citeauthoryear{Zhu, Jiang, and Xiao}{Zhu
  et~al\mbox{.}}{2011}]%
        {zhu:twinkle}
\bibfield{author}{\bibinfo{person}{Jun Zhu}, \bibinfo{person}{Zhefu Jiang},
  {and} \bibinfo{person}{Zhen Xiao}.} \bibinfo{year}{2011}\natexlab{}.
\newblock \showarticletitle{{Twinkle: A Fast Resource Provisioning Mechanism
  for Internet Services}}. In \bibinfo{booktitle}{\emph{Proceedings of the 2011
  IEEE Conference on Computer Communications (INFOCOM)}}.
  \bibinfo{pages}{802--810}.
\newblock


\end{thebibliography}

\end{document}